\documentclass[pra,twocolumn,showpacs]{revtex4}
\usepackage{epsfig}
\usepackage{graphicx}
\usepackage{amsmath}
\usepackage{natbib}
\newcommand{\bn}{\begin{eqnarray}}
\newcommand{\en}{\end{eqnarray}}
\newcommand{\eml}{\end{multline}}
\newcommand{\bml}{\begin{multline}}

\begin{document}

\title {Squeezing with classical Hamiltonians}
 \author{Tom\'{a}\v{s} Opatrn\'{y}}
 \affiliation{Optics Department, Faculty of Science, Palack\'{y} University, 17. Listopadu 12,
 77146 Olomouc, Czech Republic}

\date{\today }
\begin{abstract}
A simple formula is derived for the maximum squeezing rate which occurs at the initial stages of the squeezing process: the rate only depends on the second partial derivatives of a classical Hamiltonian.  Rules for optimum rotation of the phase space are found to keep the state optimally located and oriented for fastest squeezing. These operations transform the phase-space point of interest into a saddle point with opposite principal curvatures. Similar results are found for the Bloch-sphere phase space and spin squeezing. Application of the general formulas is illustrated by several model examples including parametric downconversion, Kerr nonlinearity, Jaynes-Cummings interaction, and spin squeezing by one-axis twisting and two-axis countertwisting. 
\end{abstract}

\maketitle

\section{Introduction}
Squeezing is an irreducible resource for quantum information processing \cite{Braunstein}. Suppressing the  noise of some physical variables in squeezed states has important applications in quantum metrology \cite{Knight,Perina,Wineland1994,Lloyd}. Sometimes it is stressed that squeezed states are purely quantum mechanical states as their Glauber $P$-representation is non-positive definite \cite{MeystreZubairy}. Therefore, to find squeezing properties of various physical systems, it is very natural to use mathematical apparatus of quantum physics.

Relations between classical and quantum predictions for various squeezing processes have been studied, discussing the similarities and differences (see, e.g., \cite{Milburn-1986,Drobny-1997,Bajer-1999,Bajer-2000}).  However, there has been no general approach showing which particular features of classical systems are responsible for noise suppression that would be analogous to squeezing production. Here, a simple formula is derived for a rate at which squeezing is generated at initial stages of the process: the maximum rate only depends on the second partial derivatives of the classical Hamiltonian. Also, simple formulas are found to determine rotations of the phase space by which one keeps the state optimally located and oriented to achieve fastest squeezing rate: they only contain first and second partial derivatives of the classical Hamiltonian. Although for precise results a full quantum calculation is necessary, the classical formulas work surprisingly well as long as the uncertainty area is not deformed significantly beyond an elliptical shape described by a variance matrix. The formulas can be used as a simple rule of thumb for squeezing prospectors who need a quick orientation in the terrain to decide where to start mining their precious resource.

A simple intuitive picture of ``classical squeezing'' is as follows. Imagine a group of tourists starting their hike in a hilly countryside. Each member of the group goes along a contour line of constant elevation, having the hill on the left and the valley on the right, with a speed proportional to the magnitude of the slope. Even though initially the group might have a circular form, moving on the uneven landscape changes the formation to be stretched in one direction and squeezed in another. Here, of course, the countryside is a phase space, elevation is the value of the Hamiltonian, and the hiking rules are the classical Hamilton equations. The group of tourists represents an ensemble of classical states, and our task is to infer from the local shape of the landscape the rate at which the group gets squeezed.

The paper is organized as follows. In Sec. \ref{SqueezingRate} the formula for the squeezing rate is derived. In Sec. \ref{Orientation} we study the question how the uncertainty area changes orientation, in Sec. \ref{Compensation} we show how to compensate the motion of the uncertainty ellipse to keep the optimum squeezing rate, and in Sec. \ref{Examples} we illustrate the general results on several examples. In Sec. \ref{Bloch} we show how the results can be generalized for spin squeezing and motion on the Bloch sphere, and we conclude in Sec. \ref{Conclusion}. Several derivations of technical nature are given in Appendixes.

\section{Squeezing rate}
\label{SqueezingRate}

Consider a classical system described by a Hamiltonian $H(x,p)$ where the quantities $x$ and $p$ are rescalled such that they have the same dimension. We consider a probability distribution $\rho(x,p)$ characterized by a variation matrix
\begin{eqnarray}
V = \left( 
\begin{array}{cc}
\langle \Delta x^2 \rangle & \langle \Delta x \Delta p \rangle \\
\langle \Delta x \Delta p \rangle  & \langle \Delta p^2 \rangle 
\end{array}
\right) \equiv \left( 
\begin{array}{cc}
V_{xx} & V_{xp} \\
V_{xp}  & V_{pp}
\end{array}
\right)
 ,
\end{eqnarray}
for a state centered in $(x_0,p_0)$.
The question is what features of the Hamiltonian determine the squeezing generation in the system.

Assume that at time $t=0$ the system is in state $(x_0+\Delta x, p_0+\Delta p)$. At short time $d t$ the system will be in a new state  $(\tilde x_0+\Delta \tilde x,\tilde p_0+\Delta\tilde p)$, where up to the first order in $d t$
\begin{eqnarray}
\tilde x_0 + \Delta \tilde x&\approx &  x_0 + \Delta  x
+ \frac{d}{dt}\left(x_0 + \Delta  x \right)
d t \nonumber  \\
 &=&  x_0 + \Delta  x + \frac{\partial H(x_0 + \Delta  x, p_0+\Delta p)}{\partial p}d t \nonumber  \\
 &\approx&  x_0 + \Delta  x +  \frac{\partial H(x_0, p_0)}{\partial p} d t \\
&& + \left( \frac{\partial^2 H(x_0, p_0)}{\partial p^2}\Delta p 
+ \frac{\partial^2 H(x_0, p_0)}{\partial x \partial p}\Delta x \right) d t \nonumber 
\end{eqnarray}
and
\begin{eqnarray}
\tilde p_0 + \Delta \tilde p&\approx &  p_0 + \Delta  p
+ \frac{d}{dt}\left(p_0 + \Delta  p \right)
d t\nonumber  \\
 &=&  p_0 + \Delta  p - \frac{\partial H(x_0 + \Delta  x, p_0+\Delta p)}{\partial x}d t\nonumber  \\
 &\approx&  p_0 + \Delta  p -  \frac{\partial H(x_0, p_0)}{\partial x} d t \\
&& - \left( \frac{\partial^2 H(x_0, p_0)}{\partial x^2}\Delta x 
+ \frac{\partial^2 H(x_0, p_0)}{\partial x \partial p}\Delta p \right) d t .\nonumber 
\end{eqnarray}
Denoting the partial derivatives at $(x_0,p_0)$ as indexes, $\partial H(x_0,p_0)/\partial x \equiv H_x$, etc.,
we can write for the new central positions in the phase space 
\begin{eqnarray}
\tilde x_0 &\approx& x_0 + H_p d t, \\
\tilde p_0 &\approx& p_0 - H_x d t, 
\end{eqnarray}
and for the new deviations
\begin{eqnarray}
\Delta \tilde x &\approx & \Delta x + \left( H_{xp} \Delta x 
+ H_{pp}\Delta p \right) d t , \\
\Delta \tilde p &\approx& \Delta p - \left( H_{xp}\Delta p 
+ H_{xx}\Delta x \right) d t .
\end{eqnarray}
Assuming $\langle \Delta x \rangle = \langle \Delta p \rangle = 0$ and expressing the new variances up to the first order in $d t$ we get
\begin{eqnarray}
 \label{delxx}
 \langle\Delta \tilde x ^2 \rangle &\approx&  \langle\Delta  x ^2 \rangle \nonumber \\
& & +2 \left(H_{xp} \langle  \Delta  x ^2 \rangle 
+ H_{pp}\langle\Delta  x \Delta p \rangle  \right)d t , \\
 \langle\Delta \tilde p ^2 \rangle &\approx&  \langle\Delta  p ^2 \rangle \nonumber \\
& & -2 \left(H_{xp} \langle\Delta  p ^2 \rangle 
+ H_{xx} \langle\Delta  x \Delta p \rangle  \right)d t
 , \\
 \langle\Delta \tilde x \Delta \tilde p \rangle &\approx&  \langle\Delta  x \Delta p \rangle 
\nonumber \\
& & +\left(H_{pp}\langle\Delta p ^2 \rangle  -
H_{xx}\langle\Delta x ^2 \rangle \right)d t .
\label{delxp}
\end{eqnarray}
These results can be described as transformation of  the variation matrix according to
\begin{eqnarray}
\tilde V & = & S V S^{T},
\label{xxs}
\end{eqnarray} 
where
\begin{eqnarray}
S = \left( \begin{array}{cc} 
1+H_{xp}d t & H_{pp}d t \\
-H_{xx}d t & 1-H_{xp}d t
\end{array}
\right)
\label{xxs1}
\end{eqnarray} 
with all terms taken up to the first order in $d t$.

The results are simple for initially isotropic and uncorrelated fluctuations, i.e., $\langle\Delta x ^2 \rangle = \langle\Delta p ^2 \rangle = \sigma^2$, and $\langle\Delta  x \Delta p \rangle =0$, where we get
\begin{eqnarray}
 \langle\Delta \tilde x ^2 \rangle &\approx&  \sigma^2 \left( 1 + 2 H_{xp}\right)d t , \\
 \langle\Delta \tilde p ^2 \rangle &\approx&  \sigma^2 \left( 1 - 2 H_{xp}\right)d t , \\
 \langle\Delta \tilde x \Delta \tilde p \rangle &\approx&  \sigma^2 \left( H_{pp} -  H_{xx}\right)d t .
\end{eqnarray}
To find the rate of squeezing generation, we express the eigenvalues of the new variance matrix $\tilde V$ as
\begin{eqnarray}
\tilde V_{\pm} & = & \frac{\langle\Delta \tilde x ^2 \rangle
+ \langle\Delta \tilde p ^2 \rangle}{2} \nonumber \\
& & \pm \frac{1}{2}\sqrt{\left(\langle\Delta \tilde x ^2 \rangle- \langle\Delta \tilde p ^2 \rangle \right)^2
+ 4 \langle\Delta \tilde x \Delta \tilde p \rangle ^2},
\end{eqnarray}
finding
\begin{eqnarray}
\tilde V_{\pm} = & \sigma^2 \left(1
\pm Q d t \right)
\end{eqnarray}
where
\begin{eqnarray}
Q & = & \sqrt{\left( H_{pp} - H_{xx}  \right)^2
+ 4 H_{xp}^2}
\label{eqQ}
\end{eqnarray}
is the squeezing rate.
Note that this formula is invariant with respect to rotations of the phase space. In points of zero gradient, $H_x=H_p=0$, it has the following geometric interpretation. If $H$ is taken in the same units as $x$ and $p$, then the principal curvatures of its graph are $\frac{1}{2}(H_{xx}+H_{pp})\pm \frac{1}{2}\sqrt{\left( H_{pp} - H_{xx}  \right)^2
+ 4 H_{xp}^2}$. Thus, in this case, $Q$ is proportional to the difference of principal curvatures of the Hamiltonian graph.

\begin{figure}
\centerline{\epsfig{file=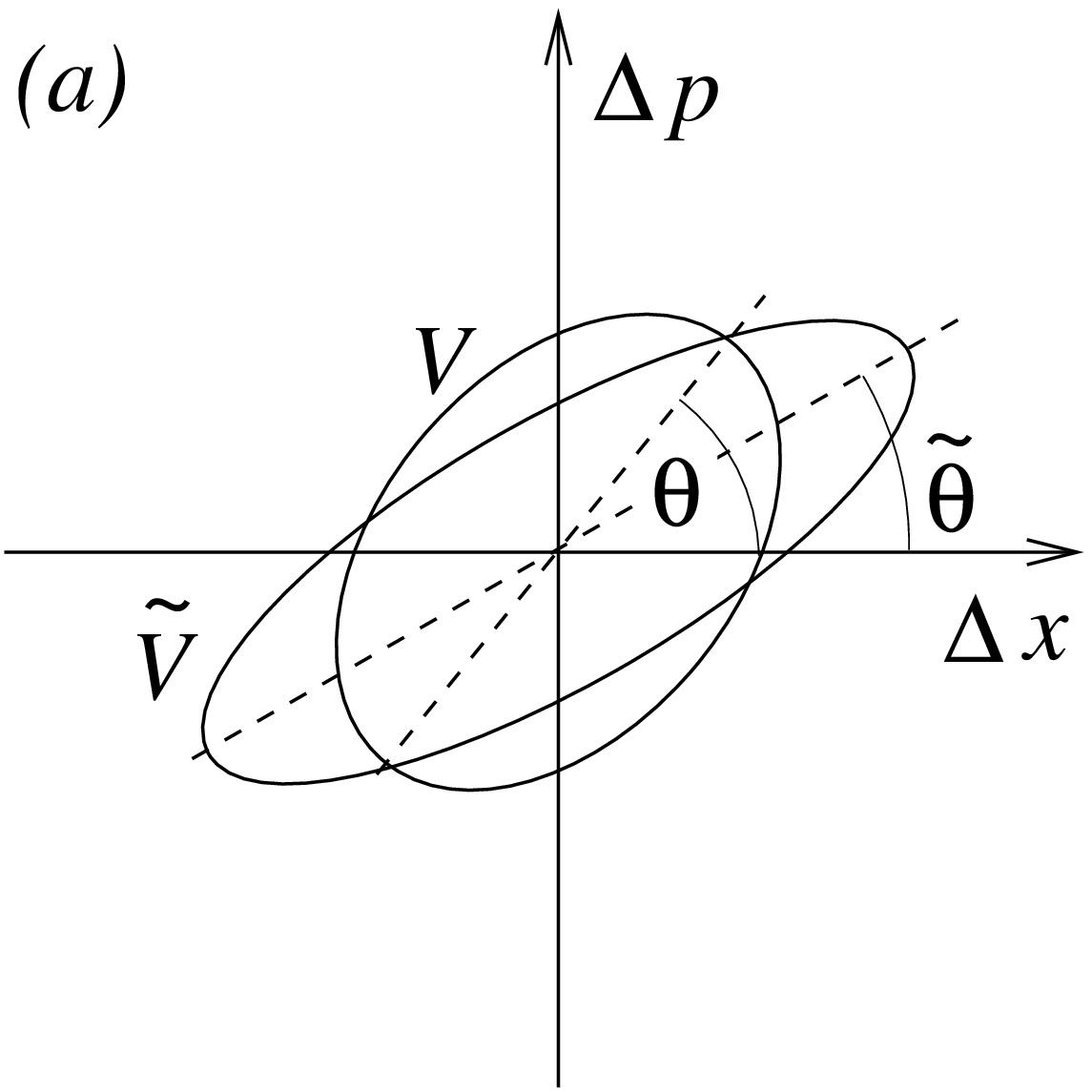,scale=0.35}
\epsfig{file=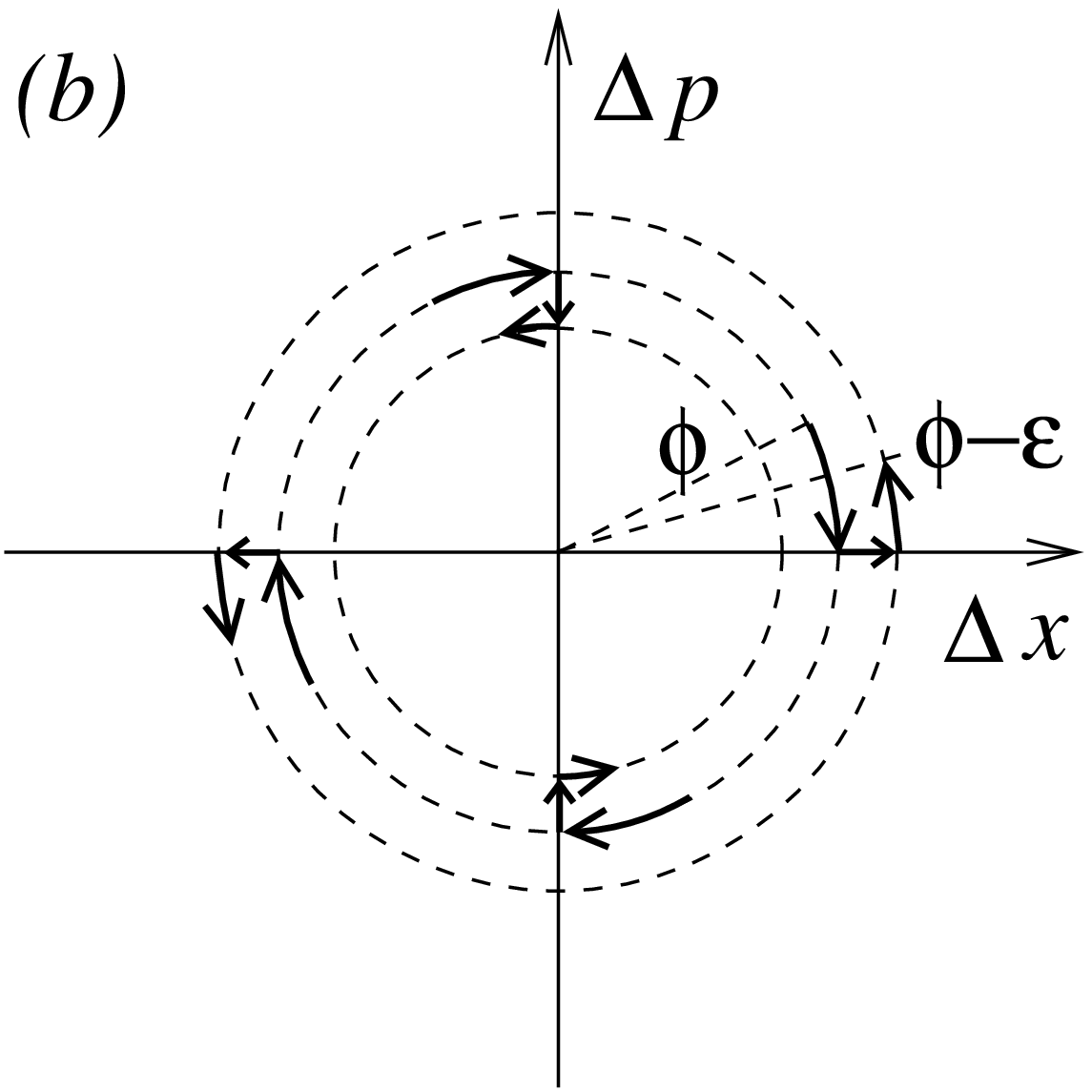,scale=0.35}}
\caption{\label{f-elipsa1}
(a) Uncertainty ellipse corresponding to variation matrix $V$ with the main axis oriented at $\theta$ is transformed into a new ellipse corresponding to  $\tilde V$ with the main axis oriented at  $\tilde \theta$.
(b) The squeezing process described by the matrix $S$ of Eq. (\ref{xxs2}) corresponds to a rotation of the phase space by $\phi$, stretching along $\Delta x$ and squeezing along $\Delta p$, and rotation of the phase space back by  $\phi-\epsilon$. Quantities $x$ and $p$ are dimensionless here and in the next figures.}
\end{figure}

\section{Orientation and rotation of the squeezing ellipse}
\label{Orientation}

As can be seen, in the special case of $H_{xx}=H_{pp}=0$ the transformation matrix $S$ of Eq. (\ref{xxs1}) is diagonal. If also $V$ is diagonal (i.e., $\Delta x$ and $\Delta p$ are uncorrelated), the transformation squeezes one of the variables and stretches the other with rate $Q$ of Eq. (\ref{eqQ}), i.e.,  $Q=2|H_{xp}|$. 
In a general case, however, the squeezing ellipse changes orientation of the main axis, as shown in Fig. \ref{f-elipsa1}a. This process can be described as follows (see Fig. \ref{f-elipsa1}b): the variation matrix is rotated by $\phi$ to a new coordinate system where $S$ is diagonal, then squeezing and stretching occurs along the new coordinates, and the variation matrix is rotated back by a modified angle $\phi-\epsilon$. Thus, the transformation matrix $S$ can be written as
\begin{eqnarray}
S &=& \left( \begin{array}{cc} 
\cos \left(\phi-\epsilon\right) & -\sin \left(\phi-\epsilon\right) \\
 \sin \left(\phi-\epsilon\right) & \cos \left(\phi-\epsilon\right) 
\end{array}
\right)
\left( \begin{array}{cc} 
1+\frac{Qd t}{2} & 0 \\
0 & 1-\frac{Qd t}{2}
\end{array}
\right) \nonumber \\
& & \times \left( \begin{array}{cc} 
\cos  \phi  & \sin  \phi  \\
 \sin  \phi  & \cos  \phi  
\end{array}
\right) .
\label{xxs2}
\end{eqnarray} 
Expanding this expression up to the first order in $\epsilon$ and $d t$, one finds
\begin{eqnarray}
S &=& \left( \begin{array}{cc} 
1+ \frac{Qd t}{2}\cos 2\phi & \epsilon + \frac{Qdt}{2}\sin 2\phi \\[2ex]
-\epsilon + \frac{Qd t}{2}\sin 2\phi  & 1- \frac{Qdt}{2}\cos 2\phi
\end{array}
\right).
\label{xxs3}
\end{eqnarray} 
Comparing this with Eq. (\ref{xxs1}) one finds
\begin{eqnarray}
Q \cos 2\phi & =& 2 H_{xp} , \\
Q \sin 2\phi & =&  H_{pp}-H_{xx}  , \\
2 \epsilon &=& \left(H_{xx}+H_{pp} \right) d t.
\end{eqnarray}
This yields
\begin{eqnarray}
\tan 2\phi &=& \frac{H_{pp}-H_{xx}}{2H_{xp}} ,
\label{EqPhi}
\end{eqnarray}
and assuming that $\epsilon$ evolves with time as $\epsilon=\omega_v  d t$, one gets 
\begin{eqnarray}
\omega_v &=& \frac{H_{xx}+H_{pp}}{2}
\label{EqOmega}
\end{eqnarray}
with $Q$ given by Eq. (\ref{eqQ}).
Eq. (\ref{EqPhi}) tells us what is the best orientation of the main axis of the uncertainty ellipse to generate squeezing the fastest way, namely, $\theta=\phi$. Eq. (\ref{EqOmega}) tells us with what rate should one rotate the system to keep the uncertainty ellipse optimally oriented, namely $\omega =-\omega_v$.

\section{Compensation of motion of the  uncertainty ellipse}
\label{Compensation}

\begin{figure}
\centerline{\epsfig{file=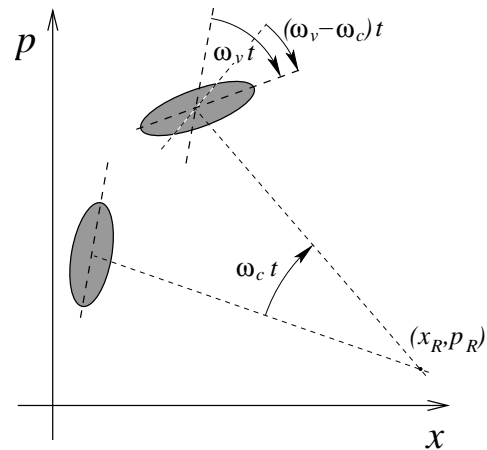,scale=0.5}}
\caption{\label{f-elipsa7}
Motion of the uncertainty ellipse as combination of rotation of its center around $(x_R,p_R)$ with rate $\omega_c$, and change of its orientation with rate $\omega_v$.
}
\end{figure}

During the evolution the uncertainty ellipse not only deforms, but also drifts through the phase space.
It is convenient to express the motion of the center of the ellipse as rotation around a phase space point $(x_R,p_R)$ with angular velocity $\omega_c$ (see Fig. \ref{f-elipsa7}). We show in Appendix \ref{ApA} that for a phase space point $(x,p)$ the rotation center is at  $(x_R,p_R)=(x,p)+(R_x,R_p)$ with
\begin{eqnarray}
 R_x &=&  -\frac{H_x}{\omega_c} ,
\label{Rx}
 \\
R_p &=&  -\frac{H_p}{\omega_c} ,
\label{Rp}
\end{eqnarray}
with the angular frequency of the motion of the center being
\begin{eqnarray}
 \omega_c = \frac{H_x^2 H_{pp}+ H_p^2 H_{xx}-2H_x H_p H_{xp}}{H_x^2 + H_p^2} .
\label{omegac}
\end{eqnarray}
While the center rotates around $(x_R,p_R)$ with  $\omega_c$, an uncertainty ellipse with optimally oriented main axis deforms and changes orientation with angular velocity $\omega_v$ (see Fig. \ref{f-elipsa7}).
These formulas could be useful if one is able to construct quadratic Hamiltonians of the form $H=\frac{1}{2}\omega [(x-x_c)^2+(p-p_c)^2]$ with variable parameters $\omega$, $x_c$ and $p_c$. This can be achieved, e.g., in quantum optical experiments where rotations around the phase-space origin corresponds to the accumulation of interferometric phase, and rotations around other points can be realized by combinations of interferometric phase shifts and displacements realized by mixing the quantum field with a strong coherent signal on an unbalanced beam splitter \cite{Vogel}.

Suppose we initiate the system in a state centered at $(x_0,p_0)$ for which $Q$ reaches the desired value. We want to keep the state centered here and also keep the uncertainty ellipse optimally oriented during the squeezing process. To compensate for the motion of the uncertainty ellipse center we first add the Hamiltonian $H_{\rm ad1}$ in the form
\begin{eqnarray}
 H_{\rm ad1} = -\frac{1}{2}\omega_c \left[ (x-x_R)^2 + (p-p_R)^2 \right]
\end{eqnarray}
with $(x_R,p_R)$ calculated according to the above formulas and Eqs. (\ref{Rx}), (\ref{Rp}) in $(x,p)=(x_0,p_0)$. Hamiltonian $H+ H_{\rm ad1}$ has zero gradient so that the uncertainty ellipse stays centered at $(x_0,p_0)$. If its main axis is at the beginning optimally oriented, it starts rotation with angular velocity $\omega_v-\omega_c$. To keep the optimal orientation, one adds another Hamiltonian $H_{\rm ad2}$ in the form
\begin{eqnarray}
 H_{\rm ad2} = -\frac{1}{2}(\omega_v-\omega_c) \left[ (x-x_0)^2 + (p-p_0)^2 \right]
\end{eqnarray}
which rotates the phase space around  $(x_0,p_0)$ with the appropriate frequency. As the result,  $(x_0,p_0)$ becomes a saddle point with principal curvatures of equal magnitude and opposite signs.
The additional Hamiltonians combine to a single quadratic Hamiltonian $H_{\rm ad} = H_{\rm ad1}+H_{\rm ad2}$ so that the system evolves under the Hamiltonian $H+H_{\rm ad}$ with
\begin{eqnarray}
 H_{\rm ad} = -\frac{1}{2}\omega_v \left[ (x-x_r)^2 + (p-p_r)^2 \right] + const.,
 \label{Had}
\end{eqnarray}
where the center is localized at 
\begin{eqnarray}
 (x_r, p_r) & =& (x_R,p_R)\nonumber \\
&& + \left( 1-\frac{\omega_c}{\omega_v}\right) (x_0-x_R, p_0-p_R) .
\label{xrpr}
\end{eqnarray}
As can be checked, the squeezing rate $Q$ is unchanged.

\section{Examples}
\label{Examples}

\subsection{Harmonic oscillator}
\begin{figure}
\centerline{\epsfig{file=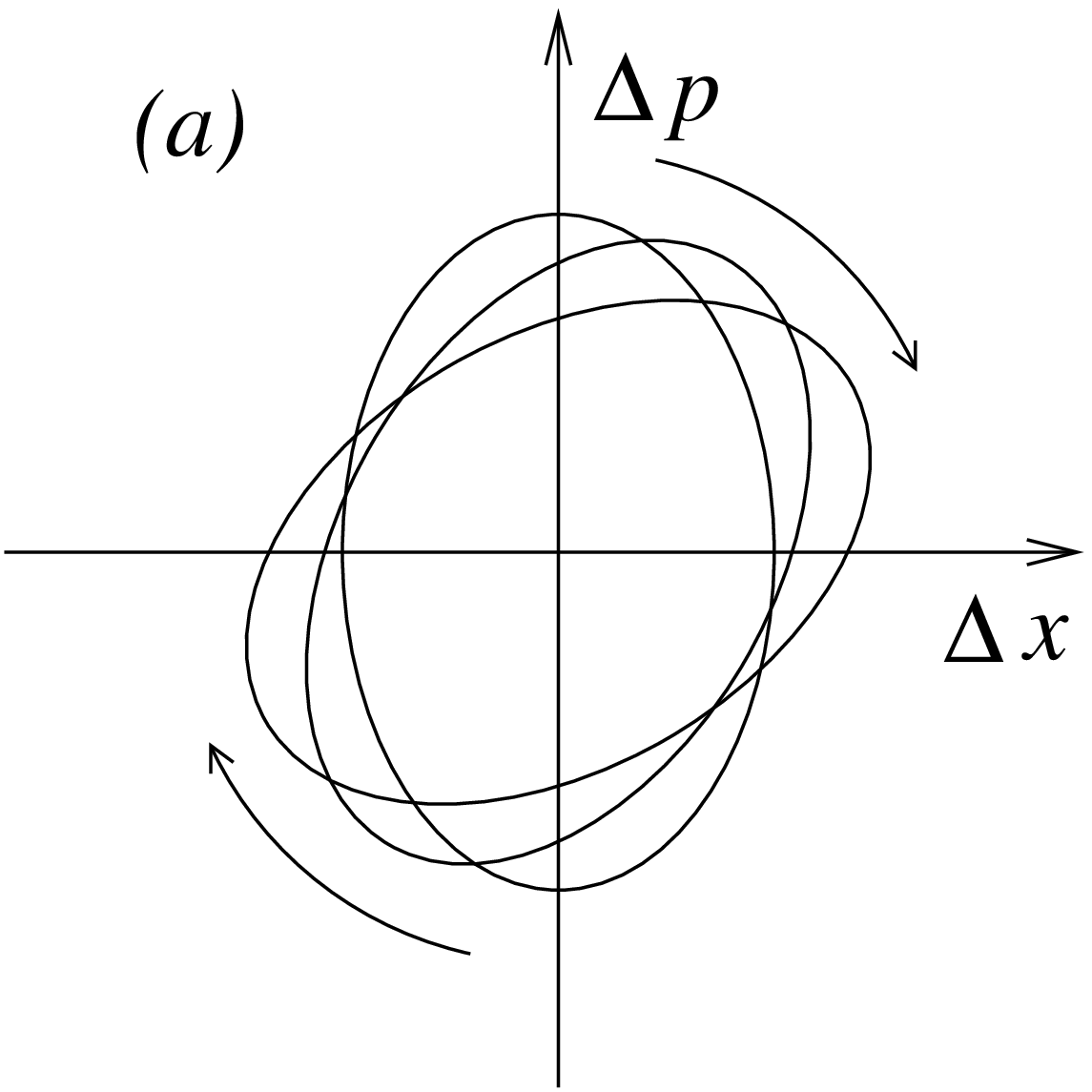,scale=0.3}
\epsfig{file=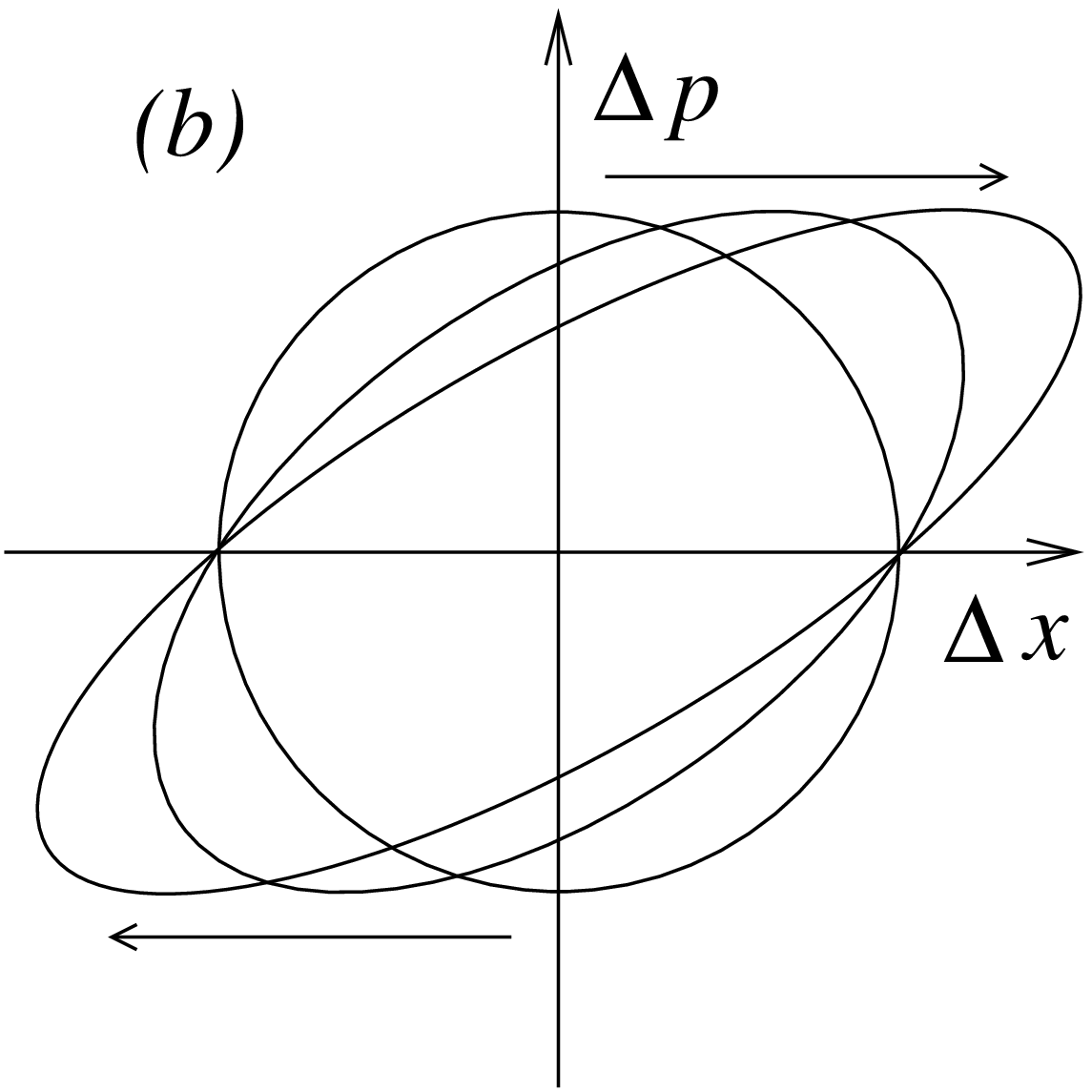,scale=0.3}}
\centerline{\epsfig{file=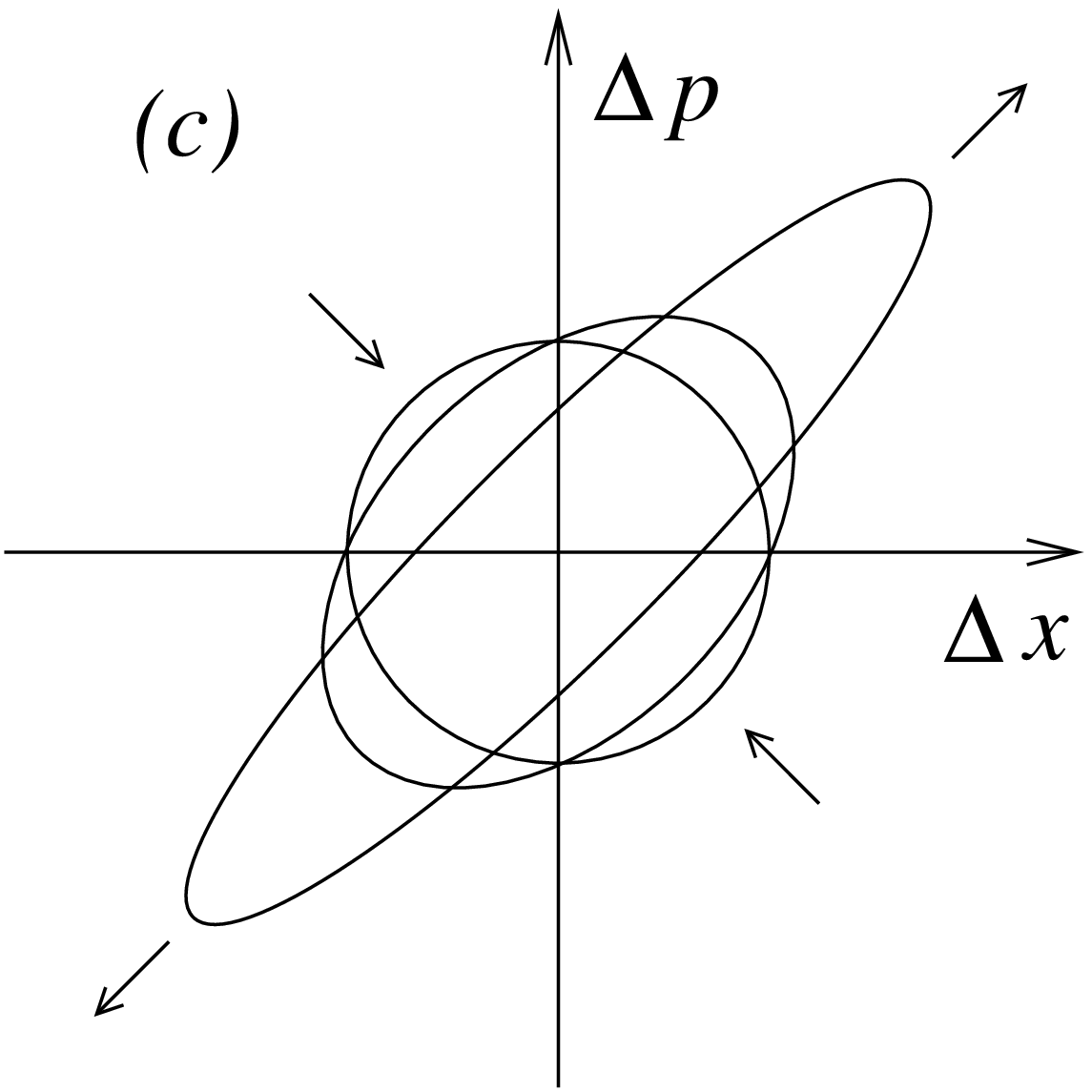,scale=0.3}
\epsfig{file=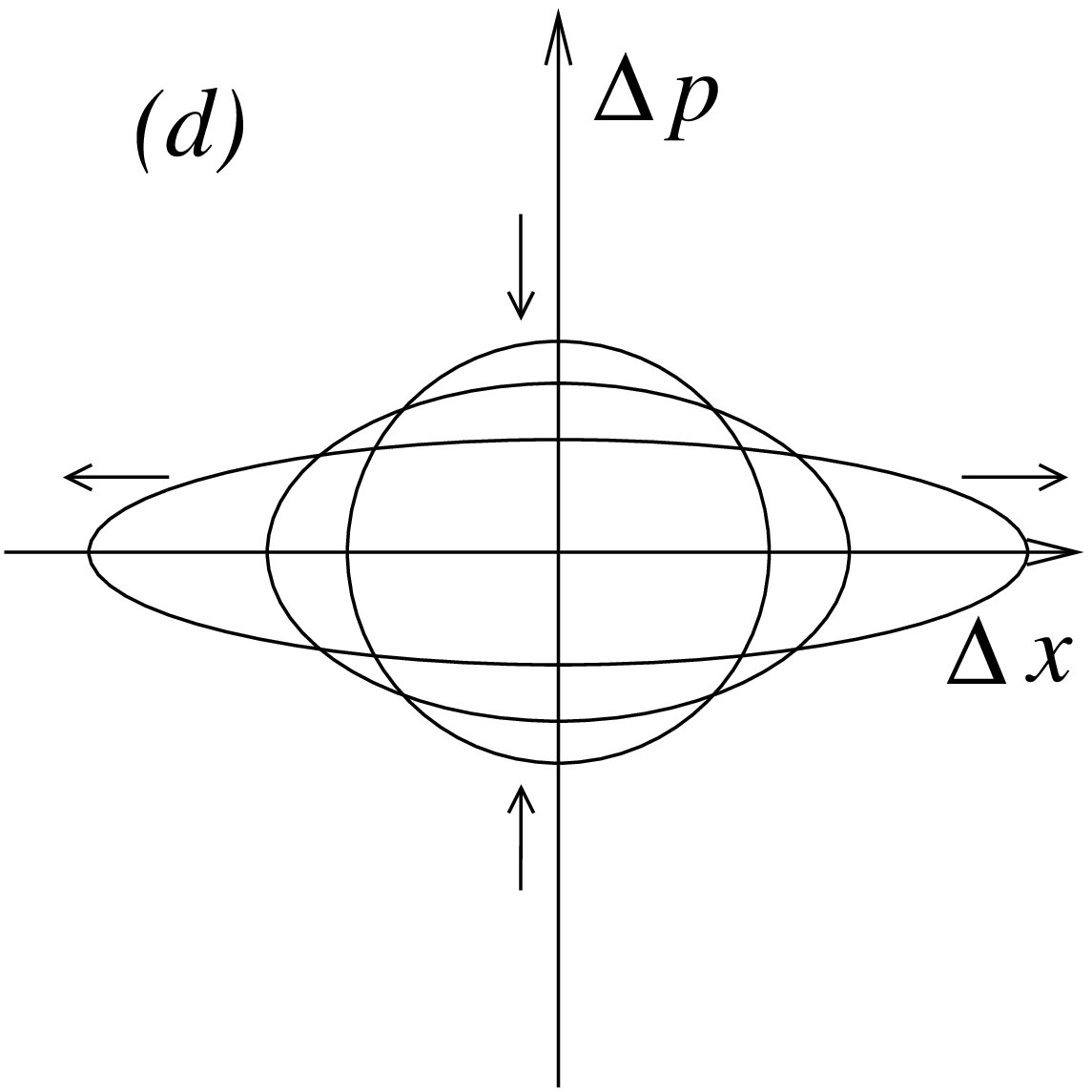,scale=0.3}}
\caption{\label{f-elipsyKvadr}
Transformation of the uncertainty ellipse for the harmonic oscillator (a), free particle (b), inverted oscillator (c), and the $xp$ Hamiltonian (d).
}
\end{figure}

The Hamiltonian is
\begin{eqnarray}
H = \frac{1}{2}\omega (p^2 + x^2),
\end{eqnarray}
and Eq. (\ref{eqQ}) yields $Q=0$, i.e., the harmonic oscillator does not produce squeezing.
Angular frequencies of Eq. (\ref{EqOmega}) and (\ref{omegac}) are $\omega_v=\omega_c=\omega$, i.e., equal to the oscillator frequency. The motion of the uncertainty ellipse is shown in Fig. \ref{f-elipsyKvadr}a.

\subsection{Quadratic Hamiltonians, parametric down conversion}

Apart from the harmonic oscillator, other Hamiltonians quadratic in $x$ and $p$ generate squeezing. In quantum optics they all have a simple interpretation of a parametric amplifier \cite{Milburn1981,Caves1981}.

\subsubsection{Free particle}
With dimensionless $x$ and $p$, the Hamiltonian of a free particle is
\begin{eqnarray}
H = \frac{1}{2m}p^2 .
\end{eqnarray}
This means that $H_{xx}=H_{xp}=0$ and $H_{pp}=1/m$
which leads to the squeezing rate $Q = 1/m$ independent of the localization in the phase space. The optimum orientation of the uncertainty ellipse is $\theta = \pi/4$ which is being rotated with the rate $\omega_v = 1/(2m)$, and $\omega_c=0$, see Fig.  \ref{f-elipsyKvadr}b. This means that to keep the optimum orientation, one has to rotate the system phase space with the rate $-1/(2m)$.

The quantum optical interpretation of this Hamiltonian is found on assuming $x$ and $p$ to be operators constructed as combinations of creation and annihilation operators, namely (assuming $m=1$)
\begin{eqnarray}
\hat x &=& \frac{1}{\sqrt{2}}\left(\hat a^{\dag} + \hat a\right), \\
\hat p &=& \frac{i}{\sqrt{2}}\left(\hat a^{\dag} - \hat a\right).
\end{eqnarray}
The Hamiltonian is then
\begin{eqnarray}
H = \frac{1}{2}\hat p^2 = -\frac{1}{4}\left(\hat a^{\dag 2} + \hat a^{2} \right)
+ \frac{1}{2}\left(\hat a^{\dag}\hat a + \frac{1}{2} \right) .
\label{Hparametric1}
\end{eqnarray}
The first term on the right hand side corresponds to a parametric down conversion with photons being created and destroyed in pairs, whereas the second term corresponds to a harmonic oscillator with frequency $\omega_v=1/2$. Using the additional Hamiltonian of Eq. (\ref{Had}) means just removing this second term.
Note that the squeezing rate $Q$ exactly corresponds to the quantum mechanical result discussed, e.g., in \cite{Knight}.

\subsubsection{Inverted oscillator}
The Hamiltonian is
\begin{eqnarray}
H = \frac{1}{2}\zeta(p^2 - x^2)
\end{eqnarray}
which leads to the squeezing rate  $Q = 2\zeta$, independent of the initial state. The optimum orientation is $\theta = \pi/4$, and the ellipse does not rotate, $\omega_v=0$. The evolution of the uncertainty ellipse is shown in Fig.  \ref{f-elipsyKvadr}c. 

In terms of quantum optical operators the Hamiltonian can be written as
\begin{eqnarray}
\hat H = \frac{1}{2}\zeta (\hat p^2 - \hat x^2) = -\frac{1}{2}\zeta \left(\hat a^{\dag 2} + \hat a^{2} \right) ,
\end{eqnarray}
corresponding to the parametric down conversion.

\subsubsection{xp-Hamiltonian}
The Hamiltonian is in the form
\begin{eqnarray}
H = \zeta xp
\label{xp}
\end{eqnarray}
which is a classical counterpart of the quantum operator
\begin{eqnarray}
\hat H &=& \frac{1}{2}\zeta \left( \hat x \hat p + \hat p \hat x \right) \nonumber \\
 &=& \frac{i}{2}\zeta \left(\hat a^{\dag 2} - \hat a^{ 2}\right)
\end{eqnarray}
corresponding to the parametric down conversion discussed in detail in \cite{Knight}. Compared to the preceding two cases it has just different phase ratio of the quadratures $x$ and $p$.
For the Hamiltonian (\ref{xp}) the squeezing rate is $Q=2\zeta$, the optimum orientation is $\theta=0$, and no rotation is generated, $\omega_v=0$. The evolution of the uncertainty ellipse is shown in Fig.  \ref{f-elipsyKvadr}d.

\subsection{Pendulum}
The Hamiltonian is
\begin{eqnarray}
H = \frac{1}{2}p^2 - \cos x
\end{eqnarray}
leading to the squeezing rate
\begin{eqnarray}
Q=1-\cos x = 2\sin ^2 \frac{x}{2},
\end{eqnarray}
which changes continuously between 0 for $x=0$ (i.e., like harmonic oscillator near the stable equilibrium) and 2 for  $x=\pi$  (i.e., like inverted oscillator near the unstable equilibrium).
The optimum orientation is $\theta =\pi/4$ and the rotation frequency is $\omega_v = \cos^2 \frac{x}{2}$ changing continuously from 1 near the stable equilibrium to 0 near the unstable equilibrium.

\begin{figure}
\centerline{\epsfig{file=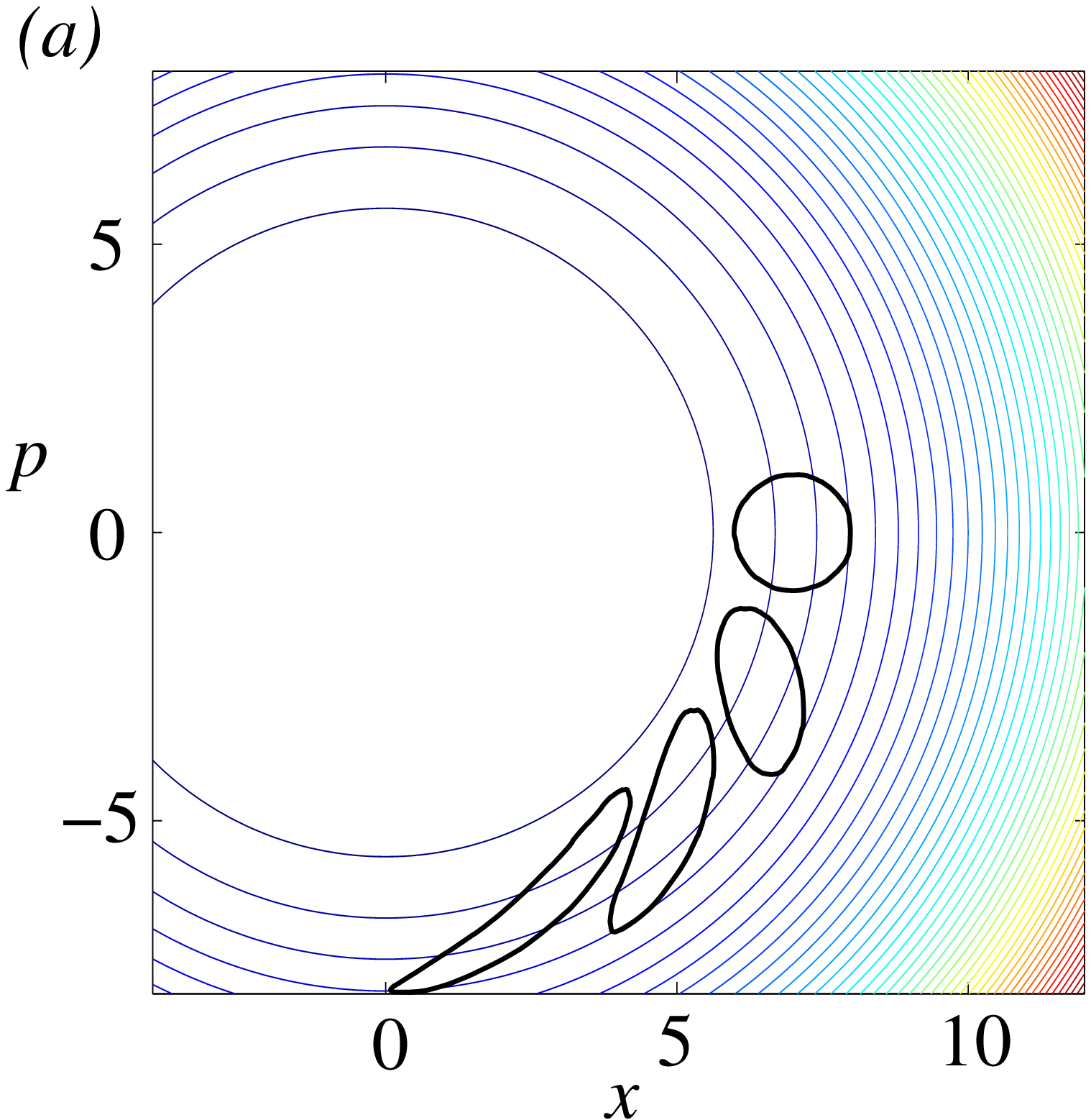,scale=0.28}
\epsfig{file=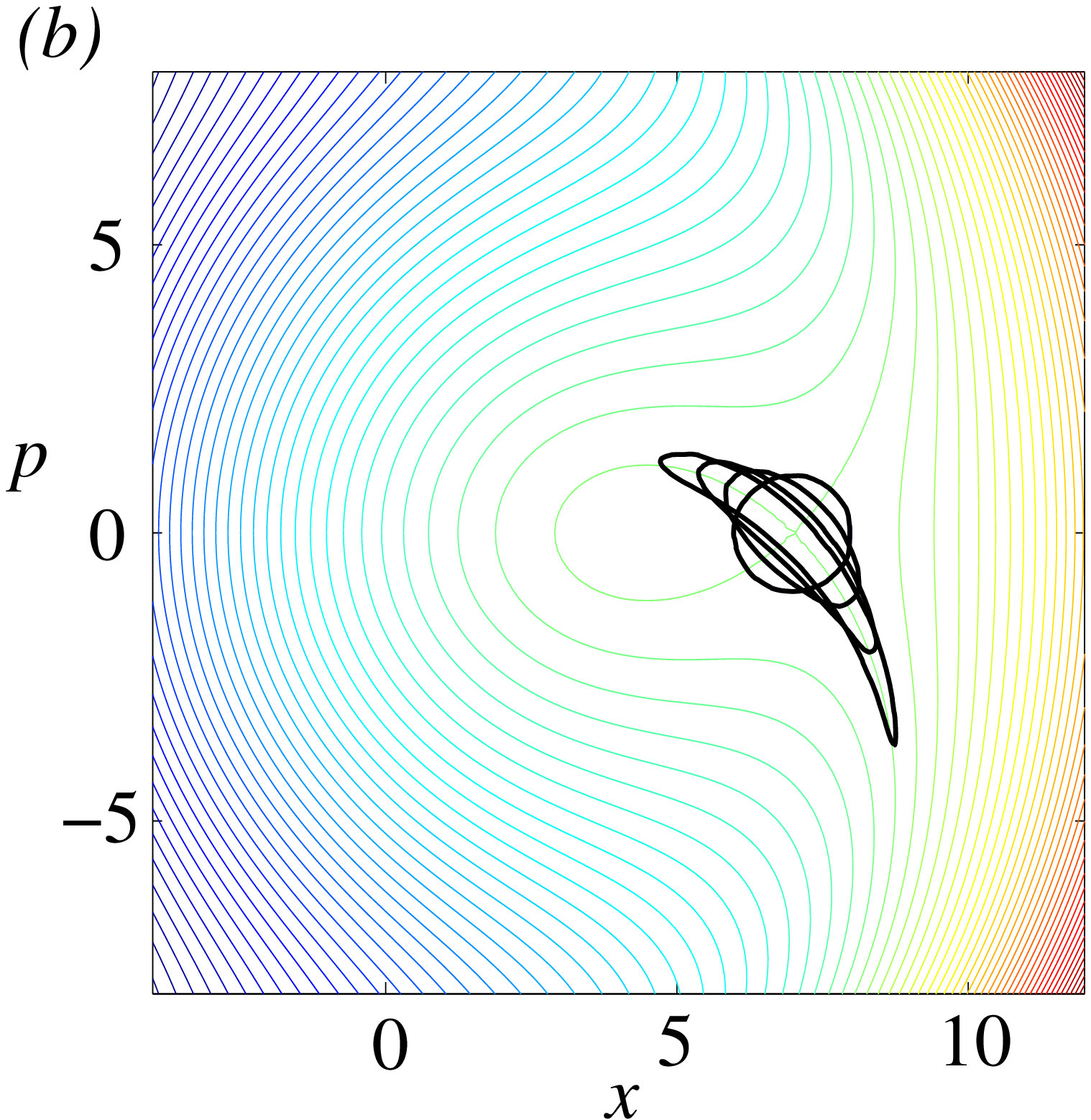,scale=0.28}}
\caption{\label{f-Kerr}
(Color online) Contour lines of the Kerr Hamiltonian and evolution of the uncertainty lines. (a) Hamiltonian (\ref{HKerr}) with $\chi=1$, (b) the same Hamiltonian with the added term (\ref{HKerrAd}). The initial state is centered at $(x_0,p_0)=(7,0)$ and the uncertainty lines are plotted in times $t=0$, $2\times 10^{-3}$, $4\times 10^{-3}$, and  $6\times 10^{-3}$.
}
\end{figure}

\subsection{Kerr nonlinearity}
Assume the Hamiltonian 
\begin{eqnarray}
H = \chi (p^2 + x^2)^2,
\label{HKerr}
\end{eqnarray}
whose quantum counterpart was shown to generate squeezing \cite{Tanas1983,Kitagawa1986}.
Eq. (\ref{HKerr}) leads to the squeezing rate  
\begin{eqnarray}
Q=8\chi (p^2 + x^2),
\label{QKerr}
\end{eqnarray}
i.e., the squeezing rate increases with the oscillator energy. Eq. (\ref{QKerr}) corresponds to the  analytical result for a quantum Kerr oscillator found in \cite{Bajer-2002}.
We show the contour lines of the Hamiltonian and the evolution of the uncertainty lines in Fig. \ref{f-Kerr}a.

The optimum orientation  depends on the phase, i.e., Eq. (\ref{EqPhi}) gives
\begin{eqnarray}
\tan 2\phi = \frac{1}{2}\left(\frac{p}{x} - \frac{x}{p} \right) .
\end{eqnarray}
Expressing the phase space point $(x,p)$ as $x=A \cos \alpha$,  $p=A \sin \alpha$ we get 
\begin{eqnarray}
\tan 2\phi = - \cot 2 \alpha,
\end{eqnarray}
so that $\phi = \alpha \pm \pi/4$ which corresponds to the optimum orientation of the axes of the uncertainty ellipse inclined by $\pm \pi/4$ from the radius. 

The angular velocities of Eqs. (\ref{EqOmega}) and (\ref{omegac}) are 
\begin{eqnarray}
\omega_v &=& 8\chi (p^2+x^2), \\
\omega_c &=& 4\chi (p^2+x^2).
\end{eqnarray}
Since $\omega_c=\omega_v/2$ and the state circles around the origin, $(x_R,p_R) = (0,0)$, from Eq. (\ref{xrpr}) we have  
\begin{eqnarray}
(x_r,p_r) = \frac{1}{2}(x_0,p_0).
\end{eqnarray}
Thus, to keep the state close to the phase state point $(x_0,p_0)$ with the uncertainty ellipse optimally oriented, one needs to use the additional Hamiltonian
\begin{eqnarray}
\label{HKerrAd}
H_{\rm ad} &=& -4\chi (x_0^2 +p_0^2) \\
&& \times
\left[ \left(x-\frac{x_0}{2} \right)^2 + \left(p-\frac{p_0}{2} \right)^2\right] .
\nonumber
\end{eqnarray}
This Hamiltonian rotates the phase space around the point in the middle between the origin and the center of the uncertainty ellipse  $(x_0,p_0)$ by twice the rate of the original rotation of  $(x_0,p_0)$ around the origin. Contour lines of the resulting Hamiltonian $H+H_{\rm ad}$ and evolution of the uncertainty lines are shown in Fig.  \ref{f-Kerr}b.

\subsection{Jaynes-Cummings model}
\label{Jaynes}

\begin{figure}
\centerline{\epsfig{file=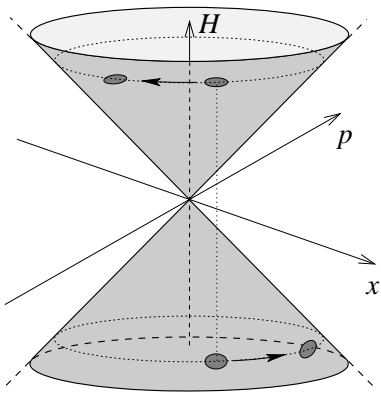,scale=0.45}}
\caption{\label{f-JC}
Hamiltonian of the Jaynes Cummings type, Eq. (\ref{HJC}) and motion of the uncertainty ellipse on the two branches.
}
\end{figure}

Assume the Hamiltonian in the form
\begin{eqnarray}
H = \pm g\sqrt{  \frac{p^2 + x^2}{2}}.
\label{HJC}
\end{eqnarray}
The motivation comes from the Jaynes-Cummings model of a two-level atom interacting with a single mode field, the quantum Hamiltonian being
\begin{eqnarray}
\hat H_{JC} = g\left(\hat a \hat \sigma_{+} + \hat a^{\dag} \hat \sigma_{-} \right),
\end{eqnarray}
where $g$ is a coupling constant and $\hat \sigma_{\pm}$ are the atomic raising and lowering operators. 
It was first shown in \cite{MeystreZubairy} that this Hamiltonian can generate squeezed states of the optical field, which was elaborated in detail in \cite{Kuklinski,Banacloche,WoodsBanacloche}.
Let us assume the initial quantum state  prepared as 
\begin{eqnarray}
|\Phi_{\pm}\rangle = |\alpha \rangle \otimes \frac{1}{\sqrt{2}}\left( |g\rangle \pm e^{i\varphi} |e\rangle \right)
\label{Phipm}
\end{eqnarray}
where $|\alpha\rangle$ is the coherent state of the field with $\alpha$ expressed as $\alpha = \sqrt{n} e^{i\varphi} = 2^{-1/2}(x+ip)$ and $|g\rangle $ and $|e\rangle $ are the ground and the excited atomic states, respectively. For times short compared to $\pi n/g$ the state remains approximately factorized so that one can study the evolution of the field separately from that of the atom. The mean energy of the state $|\Phi_{\pm}\rangle$ is $\langle \Phi_{\pm}| \hat H_{JC}|\Phi_{\pm}\rangle = \pm g\sqrt{n} = \pm 2^{-1/2}g\sqrt{x^2 + p^2}$ which corresponds to the classical Hamiltonian (\ref{HJC}).

Graph of the Hamiltonian (\ref{HJC}) is a cone shown in Fig. \ref{f-JC}, the two branches corresponding to the two signs of the atomic superposition  in Eq. (\ref{Phipm}). A state on the upper branch rotates clockwise whereas that on the lower branch counterclockwise. Note that in the quantum case, if the initial atomic state is different from $2^{-1/2}\left( |g\rangle \pm e^{i\varphi} |e\rangle \right)$, the state evolves into a superposition containing two separate coherent components of the field (i.e., a Schr\"{o}dinger cat state, see \cite{Banacloche,Buzek1992}).

\begin{figure}
\centerline{\epsfig{file=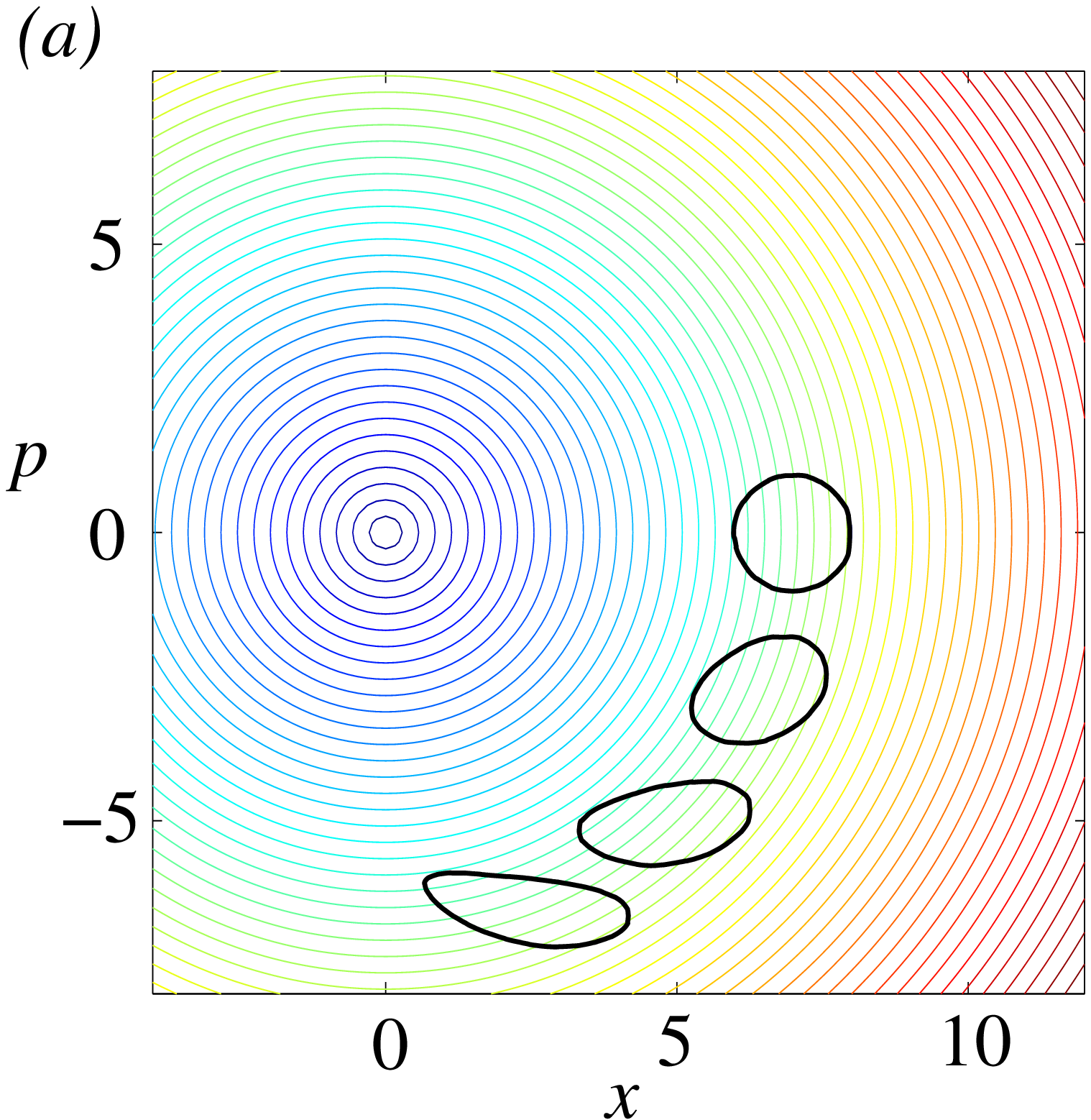,scale=0.28}
\epsfig{file=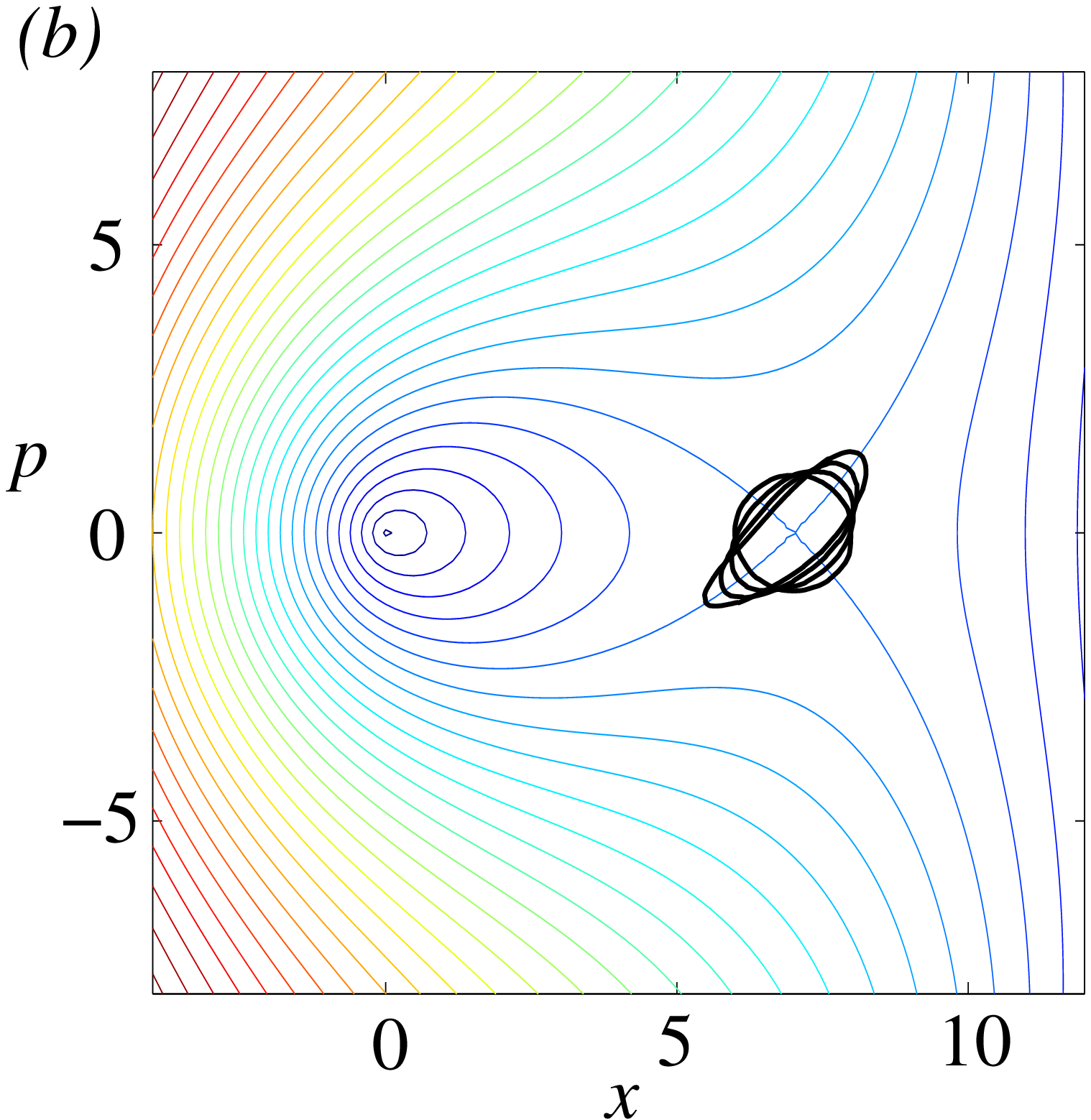,scale=0.28}}
\caption{\label{f-JCHam}
(Color online) Contour lines of the Jaynes-Cummings-like Hamiltonian and evolution of the uncertainty lines. (a) Hamiltonian (\ref{HJC}) with  the plus sign and with $g=1$, (b) the same Hamiltonian with the added term (\ref{JCHAdd}). The initial state is centered at $(x_0,p_0)=(7,0)$ and the uncertainty lines are plotted in times $t=0$, $4$, $8$, and  $16$.
}
\end{figure}

Each phase space point drifts along a circle of equal height. Since the magnitude of the cone slope does not depend on position $(x,p)$, each point moves with equal speed $\sqrt{\dot x^2 + \dot p^2} = g/\sqrt{2}$. However, points on circles of different radii move with different angular velocities: points closer to the origin describe in the same time a bigger angle than points farther from the origin. As a result, a small area of the phase space is   stretched in one direction and squeezed in the other one. We show the contour lines of Hamiltonian (\ref{HJC}) (branch with the plus sign) and evolution of the  uncertainty lines in Fig. \ref{f-JCHam}a.

On using Eq. (\ref{eqQ}) we find the squeezing rate as
\begin{eqnarray}
Q = \frac{g}{\sqrt{2}}\frac{1}{\sqrt{x^2+p^2}} .
\label{QJC}
\end{eqnarray}
Note that this rate agrees with the short-time value of squeezing evolution derived in the quantum model \cite{Banacloche}. As can be seen, contrary to the Kerr model, the squeezing rate decreases with increasing the distance from the origin. Note also, that although the classical model is equally valid for any nonzero distance from the origin, the approximation derived in the quantum model \cite{Banacloche} works only for $x^2+p^2 \gg 1$. 

The optimum orientation of the uncertainty ellipse given by Eq. (\ref{EqPhi}) is 
\begin{eqnarray}
\tan 2\phi = \frac{1}{2}\left(\frac{p}{x} - \frac{x}{p} \right) ,
\end{eqnarray}
which is the same result as in the Kerr model. However, whereas in the Kerr model the distant part of the ellipse is ahead, in the Jaynes-Cummings model the part closer to the origin is ahead.

For the angular velocities we find
\begin{eqnarray}
\omega_v = \pm \frac{g}{\sqrt{2}}\frac{1}{2\sqrt{x^2+p^2}}
\end{eqnarray}
and
\begin{eqnarray}
\omega_c = \pm \frac{g}{\sqrt{2}}\frac{1}{\sqrt{x^2+p^2}},
\end{eqnarray}
with the upper (lower) sign corresponding to the upper (lower) branch of the Hamiltonian (\ref{HJC}).
Since $\omega_c=2\omega_v$ and the state circles around the origin, $(x_R,p_R) = (0,0)$, from Eq. (\ref{xrpr}) we have  
\begin{eqnarray}
(x_r,p_r) = (-x_0,-p_0),
\end{eqnarray}
i.e., to keep the state close to the phase state point $(x_0,p_0)$ with the uncertainty ellipse optimally oriented, one needs to use the additional Hamiltonian
\begin{eqnarray}
H_{\rm ad} = \mp \frac{g}{4\sqrt{2}\sqrt{x_0^2+p_0^2}} \left[(x+x_0)^2 + (p+p_0)^2 \right].
\label{JCHAdd}
\end{eqnarray}
Contour lines of the resulting Hamiltonian $H+H_{\rm ad}$ and evolution of the uncertainty lines are shown in Fig.  \ref{f-JCHam}b.

\section{Bloch sphere as a phase space and spin squeezing}
\label{Bloch}

To describe dynamics of collective spin systems, one often depicts the states on a Bloch sphere with coordinates $J_{x}$,  $J_{y}$, and $J_{z}$ satisfying $J_x^2+J_y^2+J_z^2 = |J|^2$ where $|J|$ is a constant. These numbers are related to angular momentum operators defined as 
\begin{eqnarray}
\label{Jx}
\hat J_x &=& \frac{1}{2}(\hat{a}^{\dag} \hat{b}+ \hat{a} \hat{b}^{\dag}), \\
\hat J_y &=& \frac{-i}{2}(\hat{a}^{\dag} \hat{b}- \hat{a} \hat{b}^{\dag}), \\
\hat J_z &=& \frac{1}{2}(\hat{a}^{\dag} \hat{a}- \hat{b}^{\dag} \hat{b}), 
\label{Jz}
\end{eqnarray}
where $\hat{a}$ and $\hat{b}$ are the annihilation operators of two bosonic modes corresponding to the populations of atoms in two possible spin states. The angular momentum operators satisfy the relation $\hat J_{x}^2+\hat J_{y}^2+\hat J_{z}^2=\frac{N}{2}(\frac{N}{2}+1)$, where $N$ is the total number of particles.
The Bloch sphere has properties of a compact phase space where the classical trajectories have been used, e.g., for Bohr-Sommerfeld quantization of spin Hamiltonians \cite{Shankar1980,Garg2004}.
Here we use classical trajectories on the Bloch sphere to explore  properties of various spin squeezing models.

\subsection{Hamilton equations}
\label{BlochHamilton}

Assume Hamiltonian $H(J_{x},J_{y},J_{z})$ to be a differentiable function of $J_{x}$,  $J_{y}$, and $J_{z}$. We postulate the Hamilton equations of motion as
\begin{eqnarray}
\label{dotjx}
\dot{J_x} &=& J_{z}\frac{\partial H}{\partial J_{y}} - J_{y}\frac{\partial H}{\partial J_{z}}, \\
\dot{J_y} &=& J_{x}\frac{\partial H}{\partial J_{z}} - J_{z}\frac{\partial H}{\partial J_{x}}, \\
\dot{J_z} &=& J_{y}\frac{\partial H}{\partial J_{x}} - J_{x}\frac{\partial H}{\partial J_{y}}. 
\label{dotjz}
\end{eqnarray}
These equations can be written in a condensed form as
\begin{eqnarray}
\dot{J_i} &=& \epsilon_{ijk}J_{k}\frac{\partial H}{\partial J_{j}}
\label{dotJi}
\end{eqnarray}
where $\epsilon_{ijk}$ is the Levi-Civita symbol and Einstein summation is used, or in the vector form as
\begin{eqnarray}
\dot{\vec{J}} &=& 
{\rm grad}\ H\times \vec{J}.
\label{dotvecJ}
\end{eqnarray}
Equations (\ref{dotjx})--(\ref{dotjz}) correspond in the quantum regime to the Heisenberg equations $i\dot{\hat{A}}=[\hat{A},\hat{H}]$ where combinations of operators $\hat J_{x,y,z}$ are taken in a symmetrical form and the commutation relations are $[\hat J_{x},\hat J_{y}]= i\hat J_{z}$ with cyclical interchange of indexes. 

To see the correspondence of Eqs. (\ref{dotjx})--(\ref{dotjz}) with the classical Hamilton equations in a planar phase space, let us assume $N\gg 1$ and states near the pole with $J_z \approx N/2$ and  $J_{x,y} \ll N/2$. Defining $x\equiv \sqrt{2/N} J_x$,  $p\equiv \sqrt{2/N} J_y$, and $z\equiv \sqrt{2/N} J_z \approx \sqrt{N/2}$, we have
\begin{eqnarray}
\dot x &\approx& \frac{\partial  H}{\partial p}- p \sqrt{\frac{2}{N}} \frac{\partial H}{\partial z}, \\
\dot p &\approx& -\frac{\partial H}{\partial x}+ x \sqrt{\frac{2}{N}} \frac{\partial  H}{\partial z}.
\end{eqnarray}
For sufficiently weak dependence of $H$ on $J_z$ and large $N$, the second terms on the right hand side can be neglected and we arrive at the Hamilton equations for the 1D motion.

\subsection{Integrals of motion and the Liouville theorem}
The equations of motion guarantee that both the Hamiltonian and $J_x^2+J_y^2+J_z^2$
are conserved quantities, $dH/dt=0$ and $d(J_x^2+J_y^2+J_z^2)/dt=0$. It is obvious especially from Eq. (\ref{dotvecJ}): vector $\vec J$ moves perpendicularly to both the gradient of $H$ and to itself. Thus, the phase space point moves on a surface of a sphere along lines of constant $H$ with a speed proportional to the magnitude of ${\rm grad\ }H$.

One can also check that the Liouville theorem holds in this system. Let us assume a probability density $\rho(J_x,J_y,J_z,t)$ satisfying the continuity equation
\begin{eqnarray}
\frac{\partial \rho}{\partial t} &=& -\vec{\nabla} \cdot \vec{j},
\label{partialrho}
\end{eqnarray}
where the current density is $\vec{j} = (\dot{J_x}, \dot{J_y},\dot{J_z})\rho$.
Expressing the total time derivative of $\rho$ as
\begin{eqnarray}
\frac{d \rho}{d t} &=& \frac{\partial \rho}{\partial J_x}\dot{J_x} 
+  \frac{\partial \rho}{\partial J_y}\dot{J_y}
+  \frac{\partial \rho}{\partial J_z}\dot{J_z} 
+ \frac{\partial \rho}{\partial t}, 
\label{totalrho}
\end{eqnarray}
using Eq. (\ref{partialrho}) for $\partial \rho/\partial t$ and Eqs. (\ref{dotjx})--(\ref{dotjz}) for $\dot{J}_{x,y,z}$, one finds $d\rho/dt=0$, i.e., the Liouville theorem holds. The probability density thus behaves as an incompressible liquid circulating along constant-Hamiltonian lines on a sphere. These results hint that the Bloch sphere with equations (\ref{dotjx})--(\ref{dotjz}) represent a well-behaved phase space with classical evolution of state points.

\subsection{Evolution of moments}
Let us assume that the states are distributed in the vicinity of some phase-space point $(J_{x}^{(0)},J_{y}^{(0)},J_{z}^{(0)})$ such that the values of Hamiltonian in any nearby point  $(J_{x}^{(0)}+\Delta J_{x},J_{y}^{(0)}+\Delta J_{y},J_{z}^{(0)}+\Delta J_{z})$ can be expressed by means of the Taylor expansion up to the quadratic terms. Denoting $\bar{J_k}\equiv \langle J_k\rangle$, $V_{kl} \equiv \langle (J_k - \bar{J_k}) (J_l - \bar{J_l}) \rangle$, and the partial derivatives $H_k \equiv \partial H/\partial J_k$, etc., one finds (see Appendix \ref{ApB} for details of the derivation)
\begin{eqnarray}
\label{derbarJ}
\frac{d \bar{J_i}}{dt} &=& \epsilon_{ijk} \left( H_j  \bar{J_k} + H_{jl} V_{lk} \right) , \\
\frac{dV_{ij}}{dt} &=& H_l \left(\epsilon_{ilk} V_{jk} + \epsilon_{jlk} V_{ik} \right) \nonumber \\
 & & + H_{lp}\bar{J_k}\left(\epsilon_{ilk} V_{jp} + \epsilon_{jlk} V_{ip} \right) .
\label{derVij}
\end{eqnarray}
Note that for the special case of quadratic Hamiltonians $H=\omega_k J_k + \chi_{kl} J_k  J_l$ the results of Eqs. (\ref{derbarJ}) and (\ref{derVij}) coincide with the equations derived in \cite{Opatrny2015} for a  quantum description of spin squeezing in Gaussian approximation, and in particular for Hamiltonian  $H=-\omega J_x + \frac{\eta}{2} J_z^2$ describing a two-component Bose-Einstein condensate they coincide with the ``Bogoliubov backreaction'' equations of \cite{Vardi2001}.

\subsection{Squeezing rate and orientation of the uncertainty ellipse}
Let us first assume that the state is centered at the north pole of the Bloch sphere with $\bar{J_x}=\bar{J_y}=0$, $\bar{J_z} >0$ with no fluctuations in the radial direction, $V_{zk}\approx 0$, $k=x,y,z$.
We express the variation matrix  by means of the principal variances $V_{\pm}$, where
\begin{eqnarray}
V_{\pm}= \frac{V_{xx}+V_{yy}}{2}
\pm \frac{1}{2}\sqrt{(V_{xx}-V_{yy})^2+4V_{xy}^2},
\end{eqnarray}
and
\begin{eqnarray}
V_{xx}&=&  V_+ \cos^2 \alpha+  V_-\sin^2 \alpha, \\
V_{yy}&=&V_+ \sin^2 \alpha  + V_- \cos^2 \alpha, \\
V_{xy}&=& \frac{V_+-V_-}{2}\sin 2\alpha ,
\end{eqnarray}
where $\alpha$ is the orientation of the uncertainty ellipse.
Expressing the time derivatives by means of Eq. (\ref{derVij}) as
 \begin{eqnarray}
\label{dotVxxBloch}
\dot V_{xx}&=&  -2H_z V_{xy}+2J_z(H_{xy}V_{xx}+H_{yy}V_{xy}), \\
\label{dotVyyBloch}
\dot V_{yy}&=&  2H_z V_{xy}-2J_z(H_{xy}V_{yy}+H_{xx}V_{xy}), \\
\dot V_{xy}&=&  H_z (V_{xx}-V_{yy}) +J_z (H_{yy}V_{yy}-H_{xx}V_{xx}) ,
\end{eqnarray}
we find for the principal moments
 \begin{eqnarray}
\dot V_{\pm}&=& \pm J_z \left[  (H_{yy}-H_{xx})\sin 2\alpha + 2H_{xy}\cos  2\alpha\right]V_{\pm},
\nonumber \\
& &
\end{eqnarray}
which shows the dependence of the squeezing rate on the orientation. The optimum orientation occurs for 
\begin{eqnarray}
\tan 2\alpha = \frac{H_{yy}-H_{xx}}{2H_{xy}},
\end{eqnarray}
for which we get $\dot V_{\pm}= \pm QV_{\pm}$ with
\begin{eqnarray}
Q=|\bar{J_z}|\sqrt{(H_{xx}-H_{yy})^2+4H_{xy}^2},
\label{QBloch}
\end{eqnarray}
which is analogous to Eq. (\ref{eqQ}).
Note that if the coordinates are chosen such that $H_{xy}=0$, the orientation of optimum squeezing corresponds to $\alpha=\pm \pi/4$.

In the case of a general position on the Bloch sphere  one can proceed by first transforming the coordinate system to place the state to the pole and then use Eq. (\ref{QBloch}). Expressing the general result, one finds after some algebra 
\begin{eqnarray}
Q = \sqrt{ {\rm Tr} \left( {\cal J} H^{\prime \prime} 
 {\cal J} H^{\prime \prime} \right)
+ \frac{ {\rm Tr} \left( {\cal J}^2 H^{\prime \prime} 
 {\cal J}^2 H^{\prime \prime} \right)}{|J|^2} },
\label{Q2BlochGeneral}
\end{eqnarray}
where $H^{\prime \prime}$ is the matrix of the Hamiltonian second derivatives
\begin{eqnarray}
H^{\prime \prime}  &=& \left( 
\begin{array}{ccc}
H_{xx} & H_{xy} & H_{xz} \\
H_{xy} & H_{yy} &H_{yz}   \\
H_{xz} & H_{yz} &H_{zz} 
\end{array}
\right) ,
\end{eqnarray}
${\cal J}$ is the antisymmetric matrix corresponding to the coordinate vector $\vec{J}$
as
\begin{eqnarray}
{\cal J} &=& \left( 
\begin{array}{ccc}
0 & J_z & -J_y \\
-J_z & 0 & J_x \\
J_y & -J_x & 0
\end{array}
\right) ,
\end{eqnarray}
and $|J|=\sqrt{J_x^2+J_y^2+J_z^2}$ (see Appendix \ref{ApC} for more details of the derivation).

\subsection{Rotation of the Bloch sphere to keep the optimum squeezing orientation}
In general, the state driven by $H$ is not only squeezed, but it also drifts along the Bloch sphere and gets the orientation of the uncertainty area rotated. Suppose that we want to keep the state at the chosen position and have its orientation optimal for fastest squeezing. Assuming that the initial distribution is sufficiently narrow that the  terms $H_{jl}V_{lk}$ in Eq. (\ref{derbarJ}) can be neglected, 
compensation of the drift can be achieved by adding the Hamiltonian
\begin{eqnarray}
H_{\rm ad1} &=& -H_k J_k
\\ 
&=& \vec{\omega}_c \cdot \vec{J},
\end{eqnarray}
where
\begin{eqnarray}
\vec{\omega}_c &=& - {\rm grad} H ,
\end{eqnarray}
and the derivatives of $H$ are taken in $(J_x,J_y,J_z) =(\bar J_x, \bar J_y, \bar J_z)$.

To find the rotation Hamiltonian that would keep the state optimally oriented,
let us first consider that the chosen location is the north pole, i.e., $\bar{J_x}=\bar{J_y}=0$, $\bar{J_z}>0$ and the coordinate system is chosen such that $H_{xy}$ = 0. This means that the optimum angle is $\alpha=\pi/4$ which can be kept if variances of $J_x$ and $J_y$ are stretched with the same rate, i.e., we have $\dot V_{xx}=\dot V_{yy}$ in Eqs. (\ref{dotVxxBloch}) and (\ref{dotVyyBloch}). This can be achieved if the Hamiltonian $H+H_{\rm ad1}$ is supplemented with another term
\begin{eqnarray}
H_{\rm ad2}=\omega_{vz}J_z 
\end{eqnarray}
with 
\begin{eqnarray}
\omega_{vz} = \frac{\bar J_z}{2}\left( H_{xx}+H_{yy} \right) .
\end{eqnarray}
This result is analogous to Eq. (\ref{EqOmega}) in the planar phase space.

For a general position $(\bar J_x, \bar J_y, \bar J_z)$ on the Bloch sphere one can transform the coordinate system to get 
\begin{eqnarray}
H_{\rm ad2} &=& \frac{1}{2}\left( H_{kk} - \frac{\bar J_i H_{ij} \bar J_j}{| \bar J|^2} 
 \right) \bar J_l J_l \\
&=& \vec{\omega}_v\cdot \vec{J},
\end{eqnarray}
where 
\begin{eqnarray}
\vec{\omega}_v= \frac{1}{2}\left( {\rm Tr} H^{\prime \prime}
- \frac{ \vec{\bar J} H^{\prime \prime} \vec{\bar J}}{ | \bar J|^2}
 \right) \vec{\bar J}  ,
\end{eqnarray}
and the derivatives of $H$ are taken in $\vec{J} = \vec{\bar J}$.

\begin{figure}
\centerline{\epsfig{file=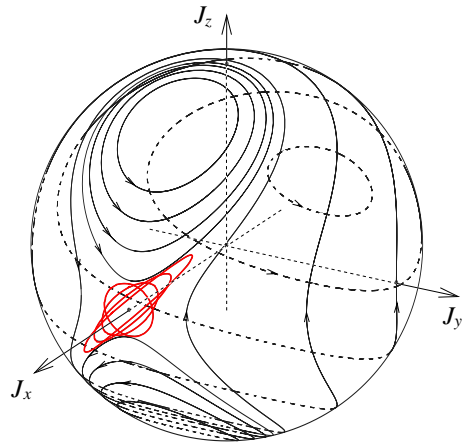,scale=0.32}}
\caption{\label{f-kouleOAT1}
(Color online) Phase space trajectories and evolution of the uncertainty region in the one-axis twisting scenario with Hamiltonian (\ref{HtotOAT}). Quantities $J_{x,y,z}$ are dimensionless here and in the next figures.
}
\end{figure}

\subsection{Examples}
\subsubsection{One-axis twisting}
The simplest Hamiltonian used to generate spin squeezing is
\begin{eqnarray}
H=\chi J_z^2
\end{eqnarray}
corresponding to the one-axis twisting introduced in \cite{Kitagawa}. Applying Eq. (\ref{Q2BlochGeneral}) we find
\begin{eqnarray}
Q &=& 2\chi \frac{\bar J_x^2+\bar J_y^2}{|\bar J|} \nonumber \\
&=& 2\chi |\bar J| \sin ^2 \theta ,
\end{eqnarray}
where $\bar J_x=|\bar J|\sin \theta \cos \phi$, $\bar J_y=|\bar J|\sin \theta \sin \phi$, and $\bar J_z=|\bar J|\cos \theta$. Thus, the fastest generation of squeezing occurs on the equator of the Bloch sphere with $Q=2\chi |\bar J|$ corresponding to the quantum result $Q=N\chi$ with $N = 2 |\bar J|$ being the total particle number.
To compensate for the drift one rotates the Bloch sphere with $\vec{\omega}_c$,
\begin{eqnarray}
\vec{\omega}_c &=& 2\chi \left( 0, 0, \bar J_z \right),
\end{eqnarray}
and to keep the optimum orientation with $\vec{\omega}_v$,
\begin{eqnarray}
\vec{\omega}_v &=& \chi \left( 1 - \frac{\bar J_z^2}{|\bar J|^2} \right) \vec{ \bar J} \\
&=& \chi \sin ^2 \theta  \vec{ \bar J}.
\end{eqnarray}
Assume now an optimally located state at the equator, say, with $\bar J_y=\bar J_z =0$ and $\bar J_x >0$. In this case there is no drift to compensate ($\omega_c=0$), and the optimal orientation of the uncertainty ellipse is kept by rotation with $\vec{\omega}_v=\chi(|\bar J|,0,0)$ so that the total Hamiltonian is
\begin{eqnarray}
H_{\rm tot} = H+H_{\rm ad2} = \chi \left( J_z^2 +|\bar J| J_x \right) .
\label{HtotOAT}
\end{eqnarray}
This Hamiltonian occurs in processes of classical bifurcation studied, e.g., in \cite{Zibold2010}. We show the corresponding classical trajectories and the evolution of the uncertainty area in Fig. \ref{f-kouleOAT1}.
These trajectories are equivalent to those found in the $\theta-\phi$ plane by semiclassical analysis of the model in \cite{Micheli2003}.

\subsubsection{Two-axis countertwisting}
Assume the Hamiltonian
\begin{eqnarray}
H=\chi \left( J_x^2 - J_y^2\right)
\label{HTACT}
\end{eqnarray}
whose quantum counterpart corresponds to the two-axis countertwisting of \cite{Kitagawa}. Possible physical realization of such a Hamiltonian has been studied recently in \cite{Opatrny2015,OKD14}.

For the squeezing rate we find
\begin{eqnarray}
Q &=& \frac{\chi}{|\bar J|}\sqrt{(\bar J_x^2-\bar J_y^2)^2+4\bar J_z^2 |\bar J|^2} \\
&=& \chi |\bar J|\sqrt{\sin ^4 \theta \cos^2 2\phi + 4 \cos^2 \theta} ,
\end{eqnarray}
which is maximized at the poles $\bar J_x = \bar J_y = 0$ with $Q=2 \chi |\bar J|$ and is zero at four points at the equator, $\bar J_z=0$ and $\phi = \pm \pi/4$, $\pi\pm \pi/4$.

\begin{figure}
\centerline{\epsfig{file=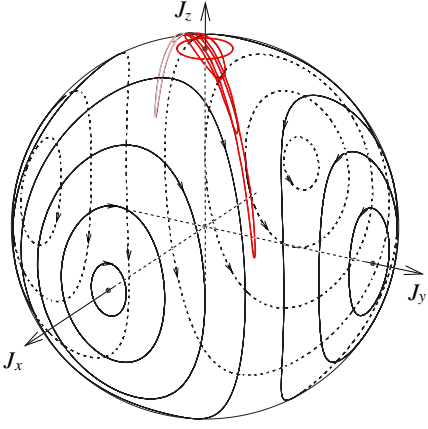,scale=0.3}}
\caption{\label{f-kouleTACT1}
(Color online) Phase space trajectories and evolution of the uncertainty region in the two-axis contertwisting scenario with Hamiltonian (\ref{HTACT}).
}
\end{figure}

To compensate for the drift one needs to rotate the Bloch sphere with the angular velocity
\begin{eqnarray}
\vec \omega_c = 2\chi (-\bar J_x, \bar J_y, 0),
\end{eqnarray}
and to keep the optimal orientation the sphere is rotated with 
\begin{eqnarray}
\vec \omega_v &=& -\chi \frac{\bar J_x^2 -\bar J_y^2}{|\bar J|^2} \vec{\bar J} \\
&=& -\chi \sin^2 \theta \cos 2\phi  \vec{\bar J} .
\end{eqnarray}
As can be seen, no rotation is necessary if the state is located at the optimum squeezing points  $\bar J_x = \bar J_y = 0$, where $\omega_c=\omega_v=0$. This case is illustrated in Fig. \ref{f-kouleTACT1}.

The results of the one-axis twisting and two-axis countertwisting scenarios correspond exactly to the quantum results obtained from the Gaussian approximation in \cite{Opatrny2015} using the ``twisting tensor'' approach.

\subsubsection{Spin squeezing by Jaynes-Cummings interaction}
Let us assume the Hamiltonian
\begin{eqnarray}
H=\pm g \sqrt{|\bar J|- J_z}.
\label{HJCBloch}
\end{eqnarray}
This form stems from the same considerations as in Sec. \ref{Jaynes}, assuming a two-mode field and an atom  coupled to mode $\hat b$ of the field, the atom being prepared in positive or negative superposition of the two levels. Another physical realization proposed in \cite{TOKM2012}
is a collection of atoms with states $|a\rangle$ and $|b\rangle$, the latter being coupled by a laser field to a Rydberg state $|r\rangle$ for which the Rydberg blockade prohibits more than one atom in state $|r\rangle$. On using the definitions of $x$ and $p$ of Sec. \ref{BlochHamilton}, we see that near the north pole of the Bloch sphere with $|J_{x,y}|\ll |\bar J| =N/2$, $J_z \approx |\bar J|$  Eq. (\ref{HJCBloch}) reduces to Eq. (\ref{HJC}).

\begin{figure}
\centerline{\epsfig{file=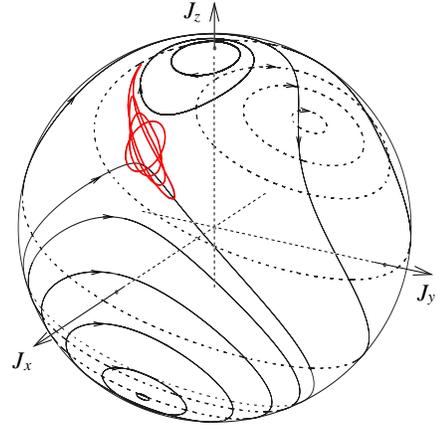,scale=0.3}}
\caption{\label{f-kouleJC}
(Color online) Phase space trajectories and evolution of the uncertainty region in the Jaynes-Cummings model with Hamiltonian $H+(\vec \omega_c+\vec \omega_v)\cdot \vec J$ of Eqs. (\ref{HJCBloch}), (\ref{omegacJC}), and (\ref{omegavJC}), with the upper choice of sign.
}
\end{figure}

For a general position, on using Eq. (\ref{Q2BlochGeneral}) one finds
\begin{eqnarray}
Q= \frac{g(|\bar J|+ J_z)}{4|\bar J| \sqrt{|\bar J|- J_z}},
\end{eqnarray}
which near the north pole of the Bloch sphere reduces to Eq. (\ref{QJC}).
For the drift compensation and for keeping optimum orientation of the uncertainty ellipse we find the angular velocities
\begin{eqnarray}
\label{omegacJC}
\vec \omega_c &=& \left(0, 0, \pm \frac{g}{2\sqrt{|\bar J|- \bar J_z}} \right) , \\
\vec \omega_v &=& \mp \frac{g}{8|\bar J|^2}\frac{|\bar J|+ \bar J_z}{\sqrt{|\bar J|- \bar J_z}} \vec{J}.
\label{omegavJC}
\end{eqnarray}
We illustrate the classical trajectories and the evolution of the uncertainty area in Fig. \ref{f-kouleJC}.
As can be checked, for states near the north pole of the Bloch sphere the combined rotation is
\begin{eqnarray}
\vec \omega_c + \vec \omega_v \approx   \frac{\mp g}{\sqrt{8|\bar J|(\bar J_x^2+\bar J_y^2)}}
\left(-\bar J_x,-\bar J_y , \bar J_z \right),
\end{eqnarray}
which means that the sphere is rotated around an axis intersecting the sphere oppositely to the state across the pole. This result is analogous to the rotation by Hamiltonian (\ref{JCHAdd}) in the planar phase space.

\section{Conclusion}
\label{Conclusion}

The main result of this paper are equations (\ref{eqQ}) and (\ref{Q2BlochGeneral}) giving the maximum squeezing rate $Q$ in a planar phase space and on the Bloch sphere, respectively. In the planar phase space, $Q$ is a function of second derivatives of the classical Hamiltonian which, in the zero-gradient points, is proportional to the difference of principal curvatures. On the Bloch sphere, formula  (\ref{Q2BlochGeneral}) generalizes the result found for quadratic Hamiltonians in \cite{Opatrny2015} where the maximum squeezing rate is proportional to the difference of the maximum and minimum eigenvalues of the twisting tensor. The formulas with the second derivatives can be interpreted as using a local expansion of the Hamiltonian up to quadratic terms to ``twist'' the phase space neighborhood of the considered state.

The other main results are the rotation frequencies of the phase space that keep the state at the right place and optimally oriented. They can be used as parameters of additional Hamiltonians to supplement the original Hamiltonian. These additional Hamiltonians themselves do not produce squeezing and their addition does not influence the value of $Q$. They can be treated rather as instruments that  optimize the exploitation, but do not change the amount of the resource. Their application transforms the point of interest into a saddle point with principal curvatures of equal magnitudes and opposite signs. 

It is interesting to note how several ``quantum'' results could be found purely by classical means. Apart from the squeezing rates, these are, e.g., the ``Bogoliubov backreaction'' equations (\ref{derbarJ}) and (\ref{derVij}) relevant for the description of two-component Bose-Einstein condensates. Although squeezing itself is sometimes described as a purely quantum phenomenon, we can see that it is not. What is ``quantum'' on squeezed states is rather the requirement on the minimum size of the uncertainty area, and sometimes the origin of the Hamiltonian governing the evolution (as, e.g., in the Jaynes-Cummings model). The classical results can be used for a quick estimation of the main properties of the states at the beginning of the squeezing process. At later stages of the evolution, however, the quantum nature of our world starts revealing in the interference phenomena that cannot be described by the classical means.

\acknowledgments
I am grateful to J. Bajer and M. Gajdacz for very useful comments and suggestions.


\appendix

\section{Rotation radius and angular velocity of a phase space point}
\label{ApA}

\begin{figure}
\centerline{\epsfig{file=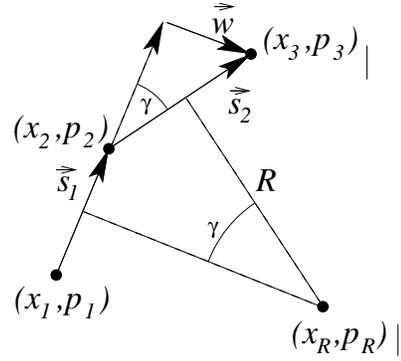,scale=0.7}}
\caption{\label{f-sipky}
Motion of the phase space point from $(x_1,p_1)$ through $(x_2,p_2)$ to $(x_3,p_3)$ expressed as rotation around $(x_R,p_R)$ by angle $\gamma$. 
}
\end{figure}

Assume a phase space point moving  from $(x_1,p_1)$ through $(x_2,p_2)$ to $(x_3,p_3)$ as in Fig. \ref{f-sipky}.  First we express $(x_2,p_2)$ and $(x_3,p_3)$ up to the second order of a short time interval $dt$.
We have
\begin{eqnarray}
x_2 &\approx & x_1 + \dot x_1 dt + \frac{1}{2}\ddot x_1 dt^2 \nonumber \\
 &=& x_1 + H_p dt + \frac{1}{2} (H_{xp}H_p-H_{pp}H_x) dt^2 , \\
p_2 &\approx & p_1 + \dot p_1 dt + \frac{1}{2}\ddot p_1 dt^2 \nonumber \\
 &=& p_1 - H_x dt + \frac{1}{2} (H_{xp}H_x-H_{xx}H_p) dt^2 , 
\end{eqnarray}
where we have used
\begin{eqnarray}
 \ddot x_1 &=& \frac{d}{dt}H_p = H_{px}\dot x_1 + H_{pp}\dot p_1 \nonumber \\
 &=& H_{xp}H_p-H_{pp}H_x, \\
 \ddot p_1 &=& -\frac{d}{dt}H_x = -H_{xx}\dot x_1 - H_{xp}\dot p_1 \nonumber \\
 &=& -H_{xx}H_p+H_{xp}H_x,
\end{eqnarray}
and the derivatives are taken in $(x_1,p_1)$. 
Similarly we have
\begin{eqnarray}
x_3 &\approx & x_2 + \dot x_2 dt + \frac{1}{2}\ddot x_2 dt^2 \\
 &=& x_2 + H_p^{(2)} dt + \frac{1}{2} (H_{xp}^{(2)}H_p^{(2)}-H_{pp}^{(2)}H_x^{(2)}) dt^2 ,\nonumber  \\
p_3 &\approx & p_2 + \dot p_2 dt + \frac{1}{2}\ddot p_2 dt^2 \\
 &=& p_2 - H_x^{(2)} dt + \frac{1}{2} (H_{xp}^{(2)}H_x^{(2)}-H_{xx}^{(2)}H_p^{(2)}) dt^2 , \nonumber 
\end{eqnarray}
where the upper index in $H_p^{(2)}$, etc., means that the derivatives are taken in  $(x_2,p_2)$. We express these derivatives in terms of the derivatives in $(x_1,p_1)$ with a correction up to the first order in $dt$ as  
\begin{eqnarray}
H_x^{(2)} &=& H_x + H_{xx}dx + H_{xp}dp \nonumber \\
&=&  H_x + (H_{xx}H_p - H_{xp}H_x) dt , \\
H_p^{(2)} &=& H_p + H_{px}dx + H_{pp}dp \nonumber \\
&=&  H_p + (H_{px}H_p - H_{pp}H_x) dt , 
\end{eqnarray}
while for the second derivatives it is enough to keep only the zeroth order of $dt$, i.e., $H_{xp}^{(2)}=H_{xp}$, etc. We thus have
\begin{eqnarray}
x_3 &=& x_1 + 2H_p dt + 2 (H_{xp}H_p-H_{pp}H_x) dt^2 , \\
p_3 &=&  p_1 - 2H_x dt + 2 (H_{xp} H_x -H_{xx} H_p ) dt^2 . 
\end{eqnarray}
We can then express the vectors $\vec{s}_1= (x_2,p_2)-(x_1,p_1)$ and  $\vec{s}_2= (x_3,p_3)-(x_2,p_2)$ of Fig. \ref{f-sipky} which are used to get their difference $\vec{w}=\vec{s}_2-\vec{s}_1$ as
\begin{eqnarray}
\vec{w}=(H_{xp}H_p-H_{pp}H_x, H_{xp} H_x -H_{xx} H_p ) dt^2 .
\end{eqnarray}
For the remaining calculations vectors $\vec{s}_{1,2}$ are taken up to the first order of $dt$ as
$\vec{s}_{1}\approx \vec{s}_{2} \equiv \vec{s}$, i.e.,  
\begin{eqnarray}
\vec{s}=(H_p,-H_x ) dt .
\end{eqnarray}
A  small angle $\gamma=\omega_c dt \ll 1$ can be expressed as 
\begin{eqnarray}
 \gamma &=& \frac{\vec{w}\times \vec{s}}{|\vec{s}|} = \frac{w_x s_p - w_p s_x}{s_x^2+s_p^2} 
\nonumber \\
&=& \frac{H_x^2 H_{pp}+ H_p^2 H_{xx}-2H_x H_p H_{xp}}{H_x^2 + H_p^2} dt ,
\end{eqnarray}
from which we get $\omega_c$ as in Eq. (\ref{omegac}).

Finally, we can express position of the center of the rotation as $(x_R,p_R) = (x_1,p_1)+(R_x,R_p)$, where
\begin{eqnarray}
 R_x &=& \frac{s_p}{\gamma},  \\
 R_p &=& -\frac{s_x}{\gamma} ,
\end{eqnarray}
from which we get Eqs. (\ref{Rx}) and (\ref{Rp}).


\section{Time derivatives of moments on the Bloch sphere}
\label{ApB}

Let us express the time derivative of $J_i+\Delta J_i$ by means of Eq. (\ref{dotJi}) as
\begin{widetext}
\begin{eqnarray}
 \frac{d}{dt}\left(J_i+\Delta J_i \right)
&=& \epsilon_{ijk} \left(J_k+\Delta J_k \right)  \frac{\partial}{\partial J_j} \left(H + H_l \Delta J_l 
+ \frac{1}{2} H_{ls} \Delta J_l \Delta J_s + \dots \right) \nonumber \\
&=& \epsilon_{ijk} J_k \left(H_j + H_{jl} \Delta J_l 
+ \frac{1}{2} H_{jls} \Delta J_l \Delta J_s + \dots \right) \nonumber \\
&& +\epsilon_{ijk} \left(H_j \Delta J_k + H_{jl} \Delta J_k \Delta J_l 
+ \frac{1}{2} H_{jls}\Delta J_k  \Delta J_l \Delta J_s + \dots \right) .
\label{BderJplusDJ}
\end{eqnarray}
If this expansion is taken in the mean value of $(J_x,J_y,J_z) = (\bar{J_x},\bar{J_y},\bar{J_z})$, one can use
$\langle \Delta J_k \rangle = 0$ and $\langle \Delta J_k \Delta J_l \rangle = V_{kl}$ so that the mean value of Eq. (\ref{BderJplusDJ}) yields

\begin{eqnarray}
 \frac{d\bar{J_i}}{dt}
&=&  \epsilon_{ijk} \bar{J_k } \left(H_j + \frac{1}{2} H_{jls} V_{ls} + \dots \right)  +\epsilon_{ijk} \left(H_{jl} V_{kl}  
+ \frac{1}{2} H_{jls}V_{kls} + \dots \right) ,
\label{BderJmean}
\end{eqnarray}
where $V_{kls}\equiv \langle \Delta J_k  \Delta J_l \Delta J_s \rangle$, etc.
If the distribution of $(J_x,J_y,J_z)$ is sufficiently narrow around its mean, we can keep in each bracket just the first term so that we arrive at Eq. (\ref{derbarJ}).

In a similar way we can express the time derivative of the product $(J_i+\Delta J_i)(J_j+\Delta J_j)$ as 
\begin{eqnarray}
 \frac{d}{dt}\left[ \left(J_i+\Delta J_i \right)(J_j+\Delta J_j) \right]
&=& \left[ \epsilon_{ilk} \left(J_j+\Delta J_j\right) + \epsilon_{jlk} \left(J_i+\Delta J_i\right)\right]
\left(J_k+\Delta J_k\right) \nonumber \\
& & \times
\left( H_l + H_{lp}\Delta J_p + \frac{1}{2}H_{lpr}\Delta J_p \Delta J_r 
+ \frac{1}{6} H_{lprs}\Delta J_p \Delta J_r \Delta J_s +  \dots \right) .
\label{BdtJiJj}
\end{eqnarray}
Expressing the mean value of Eq. (\ref{BdtJiJj}) and using (\ref{BderJmean}), one finds
\begin{eqnarray}
 \frac{dV_{ij}}{dt} &=& H_l \left( \epsilon_{ilk}V_{jk} + \epsilon_{jlk} V_{ik} \right)
+ H_{lp}  \bar{J_k }  \left( \epsilon_{ilk}V_{jp} + \epsilon_{jlk} V_{ip} \right)
\nonumber \\
&& + \frac{1}{2}H_{lpr}  \left( \epsilon_{ilk}V_{jkpr} + \epsilon_{jlk} V_{ikpr} \right) 
+ \frac{1}{6} H_{lprs} \bar{J_k }   \left( \epsilon_{ilk}V_{jprs} + \epsilon_{jlk} V_{iprs} \right) + \dots
\label{BdVijdt}
\end{eqnarray}
For a sufficiently narrow distribution only the first two terms on the right hand side of Eq. (\ref{BdVijdt})
are essential so we arrive at Eq. (\ref{derVij}).


\section{Squeezing rate for a general position on the Bloch sphere}
\label{ApC} 
For a state centered at $(\bar{J_x},\bar{J_y},\bar{J_z})$ the transformation
\begin{eqnarray}
\left( \begin{array}{c} 
J_x' \\ J_y' \\ J_z '
\end{array}
\right)
= R_2 R_1 
\left( \begin{array}{c} 
J_x \\ J_y \\ J_z 
\end{array}
\right)
\end{eqnarray}
with
\begin{eqnarray}
R_1 = \left( \begin{array}{ccc} 
\frac{\bar{J_x}}{\sqrt{\bar{J_x}^2+\bar{J_y}^2}} & \frac{\bar{J_y}}{\sqrt{\bar{J_x}^2+\bar{J_y}^2}} & 0 \\
 \frac{-\bar{J_y}}{\sqrt{\bar{J_x}^2+\bar{J_y}^2}} & \frac{\bar{J_x}}{\sqrt{\bar{J_x}^2+\bar{J_y}^2}} &  0 \\
0 & 0 & 1
\end{array}
\right) , \ \
R_2 = \left( \begin{array}{ccc} 
\frac{\bar{J_z}}{\sqrt{\bar{J_x}^2+\bar{J_y}^2 + \bar{J_z}^2}} & 0 & \frac{-\sqrt{\bar{J_x}^2+\bar{J_y}^2}}{\sqrt{\bar{J_x}^2+\bar{J_y}^2 + \bar{J_z}^2}} \\
0 & 1 & 0 \\
\frac{\sqrt{\bar{J_x}^2+\bar{J_y}^2}}{\sqrt{\bar{J_x}^2+\bar{J_y}^2 + \bar{J_z}^2}} & 0 &  \frac{\bar{J_z}}{\sqrt{\bar{J_x}^2+\bar{J_y}^2 + \bar{J_z}^2}}
\end{array}
\right) 
\end{eqnarray}
moves the state to the pole of the Bloch sphere with $\bar{J_x'}=\bar{J_y'}=0$ where one can apply Eq. (\ref{QBloch}) as $Q^2 = |J|^2\left[ (H_{x'x'}-H_{y'y'}) + 4 H_{x'y'}^2 \right]$. The derivatives with respect to the new coordinates $H_{x'x'} \equiv \partial^2 H/\partial J_x'^2$ etc are expressed using the chain rule
\begin{eqnarray}
H_{x'x'} = H_{xx}\left( \frac{\partial J_x}{\partial J_x'} \right)^2
+ 2  H_{xy} \frac{\partial J_x}{\partial J_x'} \frac{\partial J_y}{\partial J_x'}
+ \dots +  H_{zz}\left( \frac{\partial J_z}{\partial J_x'} \right)^2 ,
\end{eqnarray}
etc, with 
\begin{eqnarray}
\begin{array}{lll}
 \frac{\partial J_x}{\partial J_x'} = \frac{\bar{J_x} \bar{J_z}}{\sqrt{\bar{J_x}^2+\bar{J_y}^2}\sqrt{\bar{J_x}^2+\bar{J_y}^2 + \bar{J_z}^2}}, \quad
&  \frac{\partial J_y}{\partial J_x'} = \frac{\bar{J_y} \bar{J_z}}{\sqrt{\bar{J_x}^2+\bar{J_y}^2}\sqrt{\bar{J_x}^2+\bar{J_y}^2 + \bar{J_z}^2}}, \quad &
 \frac{\partial J_z}{\partial J_x'} = -\frac{\sqrt{\bar{J_x}^2+\bar{J_y}^2}}{\sqrt{\bar{J_x}^2+\bar{J_y}^2 + \bar{J_z}^2}}, \\
 \frac{\partial J_x}{\partial J_y'} = \frac{-\bar{J_y} }{\sqrt{\bar{J_x}^2+\bar{J_y}^2}}, \quad
&  \frac{\partial J_y}{\partial J_y'} = \frac{\bar{J_x} }{\sqrt{\bar{J_x}^2+\bar{J_y}^2}}, \quad
&  \frac{\partial J_z}{\partial J_y'} = 0, \\
 \frac{\partial J_x}{\partial J_z'} =  \frac{\bar{J_x} }{\sqrt{\bar{J_x}^2+\bar{J_y}^2 + \bar{J_z}^2}}, \quad
&  \frac{\partial J_y}{\partial J_z'} =  \frac{\bar{J_y} }{\sqrt{\bar{J_x}^2+\bar{J_y}^2 + \bar{J_z}^2}}, \quad &\frac{\partial J_z}{\partial J_z'} =  \frac{\bar{J_z} }{\sqrt{\bar{J_x}^2+\bar{J_y}^2 + \bar{J_z}^2}}.
\end{array}
\end{eqnarray}
On changing from the mean values to the coordinates of the point of interest $\bar{J_k} \to J_k$ we arrive at
\begin{eqnarray}
Q^2 &=& |J|^{-2}\left\{
\left(J_y^2+J_z^2 \right)^2 H_{xx}^2 + \left(J_x^2+J_z^2 \right)^2 H_{yy}^2
+   \left(J_x^2+J_y^2 \right)^2 H_{zz}^2 \right. \nonumber \\
& & + 4  \left(J_x^2+J_z^2 \right)\left(J_y^2+J_z^2 \right) H_{xy}^2
+ 4  \left(J_x^2+J_y^2 \right)\left(J_y^2+J_z^2 \right) H_{xz}^2
+ 4  \left(J_x^2+J_y^2 \right)\left(J_x^2+J_z^2 \right) H_{yz}^2 \nonumber \\
&& + 2\left[J_x^2J_y^2 -J_z^2 \left(J_x^2+J_y^2 +J_z^2\right) \right] H_{xx} H_{yy}
 + 2\left[J_x^2J_z^2 -J_y^2 \left(J_x^2+J_y^2 +J_z^2\right) \right] H_{xx} H_{zz}\nonumber \\
&& 
 + 2\left[J_y^2J_z^2 -J_x^2 \left(J_x^2+J_y^2 +J_z^2\right) \right] H_{yy} H_{zz}
\nonumber \\
&& -4 J_x J_y \left(J_y^2+J_z^2 \right) H_{xx}H_{xy}
-4 J_x J_z \left(J_y^2+J_z^2 \right) H_{xx}H_{xz}
+ 4 J_y J_z  \left(2 J_x^2 + J_y^2+J_z^2 \right) H_{xx}H_{yz} \nonumber \\
&& -4 J_x J_y \left(J_x^2+J_z^2 \right) H_{yy}H_{xy}
+ 4 J_x J_z  \left(J_x^2 + 2J_y^2+J_z^2 \right) H_{yy}H_{xz} 
-4 J_y J_z \left(J_x^2+J_z^2 \right) H_{yy}H_{yz}\nonumber \\
&& + 4 J_x J_y  \left(J_x^2 + J_y^2+2J_z^2 \right) H_{zz}H_{xy}  -4 J_x J_z \left(J_x^2+J_y^2 \right) H_{zz}H_{xz} -4 J_y J_z \left(J_x^2+J_y^2 \right) H_{xz}H_{yz}\nonumber \\
&& \left. -8 J_y J_z \left(J_y^2+J_z^2 \right) H_{xy}H_{xz}
-8 J_x J_z \left(J_x^2+J_z^2 \right) H_{xy}H_{yz}
-8 J_x J_y \left(J_x^2+J_y^2 \right) H_{xz}H_{yz}
\right\} ,
\end{eqnarray}
which can be abbreviated by using the Einstein summation as
\begin{eqnarray}
Q^2 &=& \frac{1}{J_wJ_w}\left[(J_sJ_s)
(\epsilon_{ijk}\epsilon_{lpq} J_k J_q H_{jl}H_{pi})
+(J_k H_{kq} J_q)  (J_s H_{sp} J_p) \right.
\nonumber \\
& & \left.
- 2 (J_kJ_k) (J_s H_{sl} H_{lq}J_q)
+ (J_kJ_k)  (J_sJ_s) (H_{lq}H_{ql})
\right] \nonumber \\
&=& \epsilon_{ijk} J_k H_{jl}\epsilon_{lpq} J_q H_{pi}
+ \frac{1}{J_wJ_w}\left(
\epsilon_{ijk} J_k \epsilon_{jlp} J_p H_{lq}
\epsilon_{qrs} J_s \epsilon_{rtv} J_p H_{ti}
\right) ,
\end{eqnarray}
which corresponds to Eq. (\ref{Q2BlochGeneral}), taking into account that ${\cal J}_{ij}=\epsilon_{ijk}J_k$.

\end{widetext}

\end{document}